\documentclass[a4paper,11pt]{article}
\usepackage{color}
\usepackage{ulem}
\usepackage{enumerate}
\usepackage{amsmath}
\usepackage{graphicx}

\def \be {\begin{equation}}
\def \ee {\end{equation}}
\def \ba {\begin{array}}
\def \ea {\end{array}}
\def \bea {\begin{eqnarray}}
\def \eea {\end{eqnarray}}

\def \ble {\begin{widetext}\begin{equation}}
\def \ele {\end{equation}\end{widetext}}
\def \blea {\begin{widetext}\begin{eqnarray}}
\def \elea {\end{eqnarray}\end{widetext}}

\def \nn {\nonumber}
\def \ketpsi {|\psi\rangle}
\def \brapsi {\langle \psi |}
\def \ketphi {|\phi\rangle}
\def  \braphi {\langle \phi|}
\def \trans {\mathcal{T}^{\psi|\phi}}
\def \transA {\mathcal{T}_A^{\psi|\phi}}
\def \transAc {\mathcal{T}_A^{\phi|\psi}}
\def \blea {\begin{widetext}\begin{eqnarray}}
\def \elea {\end{eqnarray}\end{widetext}}
\def \mO {\mathcal{O}}

\def \mA {\mathcal{A}}
\def \mX {\mathcal{X}}

\usepackage[colorlinks=true,linkcolor=red,citecolor=blue,urlcolor=blue,bookmarks]{hyperref}

\def \be {\begin{equation}}
\def \ee {\end{equation}}
\def \ba {\begin{array}}
\def \ea {\end{array}}
\def \bea {\begin{eqnarray}}
\def \eea {\end{eqnarray}}
\def \nn {\nonumber}

\def \a {\alpha}
\def \b {\beta}

\def \m {\mu}

\def \ep {\mathrm{e}}
\def \ii {\mathrm{i}}

\def \tr {\textrm{tr}}

\def \and {{~\textrm{and}~}}

\begin{document}

\title{Pseudoentropy sum rule by analytical continuation of the superposition parameter}
\author{Wu-zhong Guo$^{}$\footnote{wuzhong@hust.edu.cn}~, Yao-zong Jiang\footnote{jyz.hunan@gmail.com}~, Jin Xu\footnote{d202180103@hust.edu.cn}}


\maketitle
\date{}
\vspace{-10mm}
\begin{center}
{\it School of Physics, Huazhong University of Science and Technology,\\
 Luoyu Road 1037, Wuhan, Hubei
430074, China
\vspace{1mm}
}
\vspace{10mm}
\end{center}

\begin{abstract}
In this paper, we establish a sum rule that connects the pseudoentropy and entanglement entropy of the superposition state. Through analytical continuation of the superposition parameter, we demonstrate that the transition matrix and density matrix of the superposition state can be treated in a unified manner. Within this framework, we naturally derive sum rules for the (reduced) transition matrix, pseudo-Rényi entropy, and pseudoentropy. Furthermore, we demonstrate the close relationship between the sum rule for pseudoentropy and the singularity structure of the entropy function for the superposition state after analytical continuation. We also explore potential applications of the sum rule, including its relevance to understanding the gravity dual of non-Hermitian transition matrices and establishing upper bounds for the absolute value of pseudoentropy.
\end{abstract}

\section{Introduction}

For a given quantum system, the wave function or density matrix is expected to include all the information about the system. In finite-dimensional Hilbert spaces, one can explicitly express the wave function using a complete basis. However, in the realms of quantum field theories (QFTs) and quantum many-body systems, obtaining the exact wave function with a given basis is often infeasible. Instead, we typically study the properties of the wave function by probing the entanglement or quantum correlations among different partitions. 

In recent years, entanglement entropy (EE) has been a very useful quantity in various research fields\cite{Amico:2007ag}-\cite{Rangamani:2016dms}. Specially, in QFTs it is found to show some universal properties. In the context of AdS/CFT, EE is found to be linked to the bulk minimal surface, known as RT \cite{Ryu:2006bv} or HRT formula \cite{Hubeny:2007xt}, which gives us more insights into the nature of classical and quantum spacetime \cite{VanRaamsdonk:2010pw}-\cite{Almheiri:2019psf}. 

 It is anticipated that the RT or HRT formula should extend beyond its application solely in AdS/CFT to encompass a broader framework of gauge/gravity duality, such as dS/CFT \cite{Strominger:2001pn}\cite{Maldacena:2002vr}. An important observation is that when applying the RT or HRT formula to more general scenarios, it may yield complex-valued results on the gravity side \cite{Doi:2022iyj}. This implies that the dual quantity in boundary theories cannot be straightforwardly interpreted as EE. Thus, it suggests the necessity of generalizing the concept of EE to encompass a wider range of situations. 

A proper way to generalize EE is by transitioning from the density matrix to what is known as the transition matrix \cite{Nakata:2020luh}. In standard quantum mechanics, a system is characterized by the state  $|\psi\rangle$, or by the density matrix $\rho=|\psi\rangle\langle \psi|$. The transition matrix incorporates another state $|\phi\rangle$, thereby extending the density matrix $\rho$ to non-Hermitian $\trans=\frac{|\psi\rangle \langle \phi|}{\langle \phi|\psi\rangle}$. The non-Hermitian transition matrix exhibits very similar properties to the density matrix. If one can describe the system using the transition matrix, we can define the reduced transition matrix for subsystem $A$
\bea
\transA:= tr_{\bar A} \trans,
\eea
where $\bar A$ is the complementary part of $A$. Subsequently, we can define the quantity  $S(\transA):=-tr \transA \log \transA$ as a generalization of von Neumann entropy of $\rho_A:=tr\rho$\cite{Nakata:2020luh}, see also \cite{Murciano:2021dga}.  $S(\transA)$ is referred to as pseudoentropy (PE), and it is generally complex-valued. To compute PE we usually introduce the pseudo-R\'enyi entropy $S^{(n)}(\transA)=\frac{\log tr(\transA)^n}{1-n}$ for $n$ being integers. The PE can be obtained by analytical continuation of $n$ and taking the limit $S(\transA)=\lim_{n\to 1}S^{(n)}(\transA)$\footnote{See \cite{Mollabashi:2020yie}-\cite{He:2024jog} for recent processes.}.

A common question regarding the concept of PE is whether it can be interpreted as a type of entropy. This is actually a subtle question. On one hand, EE can be seen as a special case when $|\phi\rangle=|\psi\rangle$. On the other hand if one insist the RT or HRT formula can be applied to more general situations, it is nature to introduce the transition matrix and PE in QFTs. Furthermore, recent findings have revealed that the transition matrix is inherently linked to the density matrix of the superposition state \cite{Guo:2023aio}. To state the relation we would like to introduce a series of superposition states denoted by $|\xi(c)\rangle$, defined as
\bea\label{superpositionstate}
|\xi(c)\rangle=\mathcal{N}(c)(\ketphi+c \ketpsi),
\eea
where $\mathcal{N}(c)=1/\sqrt{1+|c|^2+c \langle \phi\ketpsi+c^* \langle \psi \ketphi}$ is the normalization constant for the state $|\xi(c)\rangle$.
We also define the reduced density matrix $\rho_A(c):=\tr_{\bar A} |\xi(c)\rangle \langle \xi(c)|$. It is expected that we can establish the following relations connecting the operator  $\transA$ and $\rho_A(c)$,
\bea\label{operatorsumrule}
 (\transA)^n=\sum_{c\in \mathcal{S}} a(c) [ \rho_A(c) ]^n ,
\eea
where $a(c)$ are n-dependent constants, the index $c$ belongs to a given set denoted by $\mathcal{S}$. 

The sum rule (\ref{operatorsumrule}) implies that the transition matrix is connected to the density matrix in a precise manner. This operator sum rule enables the derivation of a relationship between pseudo-Rényi entropy and Rényi entropy. Consequently, it offers a pathway to comprehend the physical significance of pseudo-Rényi entropy and PE.
In \cite{Guo:2023aio} a sum rule is constructed for the pseudo-R\'enyi entropy by using discrete Fourier transformation. For the sake of comparison, let's introduce the sum rule given in \cite{Guo:2023aio}. The superposition states are given by 
\bea\label{superposition}
|\xi_k\rangle=\mathcal{N}(e^{\frac{2\pi i}{2n+1}k})(\ketphi+e^{\frac{2\pi i}{2n+1}k} \ketpsi),
\eea
where $\mathcal{N}(e^{\frac{2\pi i}{2n+1}k})=1/\sqrt{2+e^{\frac{2\pi i}{2n+1}k}\langle \phi\ketpsi+e^{-\frac{2\pi i}{2n+1}k} \langle \psi \ketphi}$. The operator sum rule is 
\bea\label{opsumold}
(\transA)^n=\sum_{k=0}^{2n} a_k [\rho_A(k)]^n,
\eea
with 
\bea\label{ak}
a_k:= a(\ep^{\frac{2\pi \ii}{2n+1}k}) =\frac{\ep^{-\frac{2\pi \ii}{2n+1}kn}}{(2n+1) {{\mathcal{N}}_k^{n}}},\quad {\mathcal{N}}_k:=\mathcal{N}(\ep^{\frac{2\pi\ii}{2n+1}k})^2 \langle\phi |\psi \rangle,
\eea
where $\rho_A(k):= \rho_A(\ep^{\frac{2\pi \ii}{2n+1}k})$.

As emphasized in \cite{Guo:2023aio}, the form of the sum rule is not unique and depends on the choices of the set $\mathcal{S}$. Other forms of the sum rule can be obtained by appropriately selecting the set $\mathcal{S}$ and coefficients $a(c)$. The sum rule derived in \cite{Guo:2023aio} does not smoothly converge to the pseudoentropy in the limit as $n \to 1$. In this paper, we demonstrate that the sum rule for pseudo-R\'enyi entropy can also be expressed in integral form by using the Fourier transformation. Furthermore, we utilize this result to derive a sum rule for pseudoentropy. 

More importantly, through the construction of the new form sum rule, we discover that the transition matrix and density matrix of a superposition state can be treated in a unified manner. This is achieved by analytically continuing the real superposition parameter to complex values. In this framework, the transition matrix can be obtained through contour integration at infinity. Using the properties of complex integration, the sum rule for the transition matrix, pseudo-Rényi entropy, and PE can be derived naturally. In fact, this approach can be readily extended to other quantities defined by the transition matrix.

\section{A new form sum rule}

In this paper we would like to consider the superposition state
\bea
|\xi(\theta)\rangle= \mathcal{N}(\theta)(|\phi\rangle+ e^{i\theta}|\psi\rangle),
\eea
where $\theta\in (-\pi,\pi)$ and $\mathcal{N}(\theta)=1/\sqrt{2+e^{i\theta}\langle \phi|\psi\rangle+e^{-i\theta}\langle\psi| \phi\rangle}$ is the normalization constant. The new sum rule is expressed as
\bea\label{sumrulenew}
(\transA)^n =\frac{1}{2\pi }\int_{-\pi}^\pi d\theta e^{-i n\theta}\mathcal{N}_\theta^{n}\rho_A(\theta)^n,
\eea
where $\mathcal{N}_\theta:=\mathcal{N}(\theta)^{-2} \langle \phi|\psi\rangle^{-1}$.
The form of the above sum rule is very similar to the one (\ref{opsumold}). Here, we derive the result using Fourier transformation. Its advantages will become clear in the following sections.

Using the operator sum rule (\ref{sumrulenew}) one could establish sum rule for the pseudo-R\'enyi entropy and off-diagonal matrix elements  as we have done in \cite{Guo:2023aio}. The pseudo-R\'enyi entropy sum rule is easy to obtain by taking trace on both side of (\ref{sumrulenew}), the result is
\bea\label{pseudoRenyisumrule}
\ep^{(1-n)S^{(n)}(\transA)}=\frac{1}{2\pi}\int_{-\pi}^\pi d\theta e^{-i n\theta}\mathcal{N}_\theta^{n} \ep^{(1-n)S^{(n)}(\rho_A(\theta))}.
\eea
Consider a set of operators $\{ \mathcal{A}_j\}$ ($j=1,...,m$)  located in the subsystem $A$. If $m\le n$ we would have the following interesting relations
\bea\label{correlatorsumrule}
\prod_{j=1}^{m}\frac{\langle \phi| \mA_j|\psi\rangle}{\langle \phi| \psi\rangle}=\frac{1}{2\pi}\int_{-\pi}^\pi d\theta e^{-i n\theta}\mathcal{N}_\theta^{n} \prod_{j=1}^{m}\langle \xi(\theta)| \mA_j|\xi(\theta)\rangle.
\eea
The formula above demonstrates that the off-diagonal elements $\langle \phi| \mA| \psi\rangle$ can be related to the diagonal ones $\langle \xi(\theta)| \mA_j|\xi(\theta)\rangle$. This relation can be directly verified using the properties of Fourier transformation. In quantum field theory, it can also be derived by employing replica methods to evaluate the pseudo-R\'enyi entropy. Eq.(\ref{correlatorsumrule}) is closely connected to Eq.(\ref{pseudoRenyisumrule}) in this context.

\subsection{Proof of the operator sum rule}
Although the proof of the new operator sum rule (\ref{sumrulenew}) closely resembles the one in \cite{Guo:2023aio}, for the sake of completeness, we will briefly outline the proof here. The key step is that $\rho_A(c)$ can be expanded as follows,
\bea\label{rhoAexpand}
\rho_A(c) = \mathcal{N}(c)^2 \big( \rho_A^\phi+c \braphi \psi \rangle\transA+ c^* \brapsi \phi\rangle \transAc+c c^* \rho_A^\psi \big),
\eea
where $\rho_A^\phi:= \tr_{\bar A} \ketphi \braphi$, and $\rho_A^\psi:= \tr_{\bar A} \ketpsi \brapsi$.
The $n$-th power of $\rho_A(c)$ should include polynomial terms involving four operators on the right-hand side of (\ref{rhoAexpand}).
\bea\label{rhoAn}
&&[\rho_A(c)]^n=\sum_{\{r,s,t\}}c^{t+r} (c^*)^{s+r}\braphi \psi \rangle^t \brapsi \phi\rangle^s \mathcal{N}(c)^{2n}\{(\rho_A^\phi)^{n-r-s-t} (\transA)^t (\transAc)^s (\rho^\psi_A)^{r}\nn \\
 &&\phantom{[\rho_A(c)]^n=\sum_{\{r,s,t\}}} + \cdots \},
\eea
where $``+ \cdots "$ denotes the sum of all possible permutation terms for each fixed $r,s,t$. One of the special term in the summation is the one with $t=n,s=0,r=0$, i.e., $(\transA)^n$. One could consider the linear combination of $[\rho_A(c)]^n$ as the right hand side of (\ref{operatorsumrule}). The key point is that it is possible for only the special terms $(\transA)^n$ to remain while all others disappear. To achieve this we require the condition
\bea\label{constraint}
\sum_{c\in \mathcal{S}} a(c) c^{t+r} (c^*)^{s+r}\braphi \psi \rangle^t \brapsi \phi\rangle^s\mathcal{N}(c)^{2n}  =\delta_{s,0}\delta_{t,n},
\eea
Now we would like to consider a continuous set $\mathcal{S}=\{ e^{i\theta}\}$ with $\theta\in(-\pi,\pi)$. The summation in (\ref{constraint}) should be replaced by integration over the interval $(-\pi,\pi)$. We have the equation 
\bea\label{constraintsolve}
 \int_{-\pi}^{\pi}d\theta f(\theta)  \ep^{i(t-s-n)\theta}=\delta_{s,0}\delta_{t,n},
 \eea
where
\bea\label{fkak}
f(\theta):=a(\ep^{i\theta})\left[ e^{i\theta}\mathcal{N}(\ep^{i\theta})^2\braphi \psi\rangle\right]^n.
\eea
The solution to the above equation is $f(\theta)=\frac{1}{2\pi}$, from which we can obtain $a(e^{i\theta})$. Thus, we have established the validity of the new operator sum rule (\ref{sumrulenew}). Through a similar treatment, one could also derive an operator sum rule for the case where $|\psi\rangle$ and $|\phi\rangle$ are orthogonal.

Let us discuss why the term $(\transA)^n$ is particularly significant. The summation terms on the right hand side of (\ref{rhoAn}) actually contain many interesting terms that can be utilized to derive other measures of information. One is the term $(\rho_A^\phi)^{n-1}\rho_A^\psi$, which is employed in \cite{Lashkari:2014yva} to compute the relative entropy for the density matrices $\rho_A^\phi$ and $\rho_A^\psi$. This term corresponds to $t=s=0$ and $r=1$ in the summation in (\ref{rhoAn}). In general, we can parameterize the index $c$ as $R e^{i\theta}$, where $R\in (0,+\infty)$ and $\theta\in (0,2\pi)$. The coefficients associated with fixed $t,s,r$ in the summation of Eq.(\ref{rhoAn}) are given by
\bea
 R^{2r+t+s}e^{i(t-s)\theta} \braphi \psi \rangle^t \brapsi \phi\rangle^s \mathcal{N}(R e^{i\theta})^{2n}.
\eea
The term $(\rho_A^\phi)^{n-1}\rho_A^\psi$ corresponds to the case $t=s=0,r=1$.Its coefficient is same with the terms corresponding to $t=s=1,r=0$, that is $(\rho_A^\phi)^{n-2} \transA \transAc$ along with all their possible permutations. Consequently, isolating the term $(\rho_A^\phi)^{n-1}\rho_A^\psi$ using the method we previously discussed is not feasible.

\subsection{Analytical continuation of the superposition parameter }\label{superpositionsection}
The sum rule (\ref{sumrulenew}) can also be seen as a natural result by analytical continuation of the superposition parameter $\theta$ of the state $|\xi(\theta)\rangle$. Let us consider the unnormalized state 
\bea
|\tilde{\xi}(\theta)\rangle=|\phi\rangle+e^{i\theta}|\psi\rangle,
\eea
and the density matrix 
\bea
\tilde \rho(\theta)=|\tilde{\xi}(\theta)\rangle \langle \tilde{\xi}(\theta)|=(|\phi\rangle+e^{i\theta}|\psi\rangle)(\langle\phi|+e^{-i\theta}\langle\psi|).
\eea

If $\theta$ lies in the interval $(-\pi,\pi)$, the operator $\tilde{\rho}(\theta)$ is Hermitian. We can treat $\tilde{\rho}(\theta)$ as a function of a real parameter $\theta$. Now, we aim to perform an analytical continuation of $\theta$ to arbitrary complex values, akin to the Wick rotation method employed in QFT. The result is
\bea
\tilde{\rho}(\zeta)=(|\phi\rangle+\zeta|\psi\rangle)(\langle\phi|+\frac{1}{\zeta}\langle\psi|),
\eea
where $\zeta$ is complex. The function $\tilde{\rho}(\zeta)$ is defined on the complex plane, with the density matrix $\tilde{\rho}(\theta)$ lying specifically on the unit circle $|\zeta|=1$. It's worth noting that $\tilde{\rho}(\zeta)$ is Hermitian only when evaluated on points lying on the unit circle $|\zeta|=1$.

\begin{figure}
\centering\includegraphics[width=5cm]{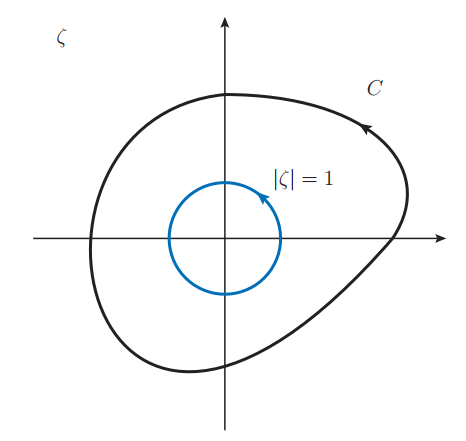}
\caption{The contours for evaluating the integral (\ref{integral11}) and (\ref{integral12}) on the complex $\zeta$ plane. The contour can be deformed to the unit circle (depicted in blue).}\label{couter}
\end{figure}

$\tilde{\rho}(\zeta)$ has simple poles at $\zeta=0$ and $\zeta=\infty$. One could extract the off-diagonal element $|\phi\rangle\langle \psi|$ by using the integral
\bea\label{integral11}
|\phi\rangle\langle \psi|=\frac{1}{2\pi i}\oint_{C}d\zeta \tilde{\rho}(\zeta),
\eea
where $C$ represents an arbitrary closed counterclockwise contour enclosing the point $z=0$, see Fig.\ref{couter}. Similarly, we have
\bea\label{integral12}
|\psi\rangle\langle \phi|=\frac{1}{2\pi i}\oint_{C}d\zeta \frac{1}{\zeta^2}\tilde{\rho}(\zeta),
\eea
where $C$ denotes the same contour shown in Fig.\ref{couter}. The contours $C$ can be deformed to the unit circle $|\zeta|=1$. One could also extract $|\psi\rangle\langle \phi|$ from the pole at infinity $\zeta=\infty$, that is
\bea
|\psi\rangle\langle \phi|=\frac{1}{2\pi i}\oint_{C}d\zeta \tilde{\rho}(1/\zeta),
\eea
where $C$ is again the contour surrounding $\zeta=0$.
we can derive the density matrix under the dephasing channel as
\bea
|\phi\rangle \langle \phi|+|\psi\rangle \langle \psi|=\frac{1}{2\pi i}\oint_{C} \zeta^{-1} \tilde{\rho}(\zeta).
\eea
 By changing the coordinate $\zeta=e^{i\theta}$ the off-diagonal elements $|\phi\rangle\langle \psi|$ and $|\psi\rangle \langle \phi|$ can be written as linear combinations of the   density matrix $\tilde{\rho}(\theta)$.

The sum rule (\ref{sumrulenew}) can be understood by similar way. The first step is to define the reduced operator $\tilde{\rho}_A(\zeta):= tr_{\bar A}\tilde{\rho}(\zeta)$. $\tilde{\rho}_A(\zeta)$ and its n-th power $(\tilde{\rho}_A(\zeta))^n$ with $n$ being integers are both meromorphic functions. By calculation we have
\bea
(\tilde{\rho}_A(\zeta))^n=\zeta^n (\tilde{\mathcal{T}}_A^{\psi|\phi})^n+...,
\eea
where $\tilde{\mathcal{T}}_A^{\psi|\phi}:= tr_{\bar A}|\psi\rangle \langle \phi|$,  ``+...'' represents terms involving powers of 
$z$ less than $n$. Thus we obtain
\bea
(\tilde{\mathcal{T}}_A^{\psi|\phi})^n=\frac{1}{2\pi i} \oint_{C} \zeta^{-n-1} (\tilde{\rho}_A(\zeta))^n d\zeta. 
\eea
Specially, the contour $C$ can be chosen as the unit circle $|\zeta|=1$. This choice results in the operator sum rule (\ref{sumrulenew}) upon restoring the normalization constants for $\tilde{\mathcal{T}}_A^{\psi|\phi}$ and $\tilde{\rho}_A(\zeta)$. Define $\rho(\zeta)=\mathcal{N}(\zeta)\tilde\rho (\zeta)$, with $\mathcal{N}(\zeta)=\zeta (2\zeta+\zeta^2 \langle \phi| \psi\rangle +\langle \psi|\phi\rangle)^{-1}$,  we have
\bea
(\transA)^n=\frac{1}{2\pi i} \oint_{C} \zeta^{-n-1}\mathcal{N}_\zeta^n \rho(\zeta)^nd\zeta,
\eea
where $\mathcal{N}_\zeta=\mathcal{N}(\zeta)^{-1}\langle \phi| \psi\rangle^{-1}$. Taking $\zeta=e^{i\theta}$ we obtain the sum rule (\ref{sumrulenew}).

The above discussion can be extended to more general cases. We begin with the function $\tilde{\rho}(\zeta)$ which has simple poles on the complex $\zeta$ plane. The function $(\tilde{\rho}_A(\zeta))^n$ stands out due to the specific operations of partial trace and taking the $n$-th power. Indeed, it remains a meromorphic function, with its poles clearly identifiable. In general, the inclusion of more generalized quantum channel operations $\mathcal{R}(\tilde{\rho})$ may lead to result functions that are not meromorphic. These functions could possess more intricate singularity structures, such as branch cuts. In such cases, careful consideration of the chosen contours becomes imperative. In the following we will show the entanglement entropy as examples.

The analytical continuation of the superposition parameter can indeed be extended to multiple superposition states. For instance,
\bea
\tilde{\rho}(\theta_1,...,\theta_m):= \sum_{k=1}^m\sum_{k'=1}^m e^{i \theta_k}|\psi_k \rangle \langle \psi_{k'}|e^{-i\theta_{k'}}
\eea
By extending the real parameters $\{\theta_k\}$ ($k=1,...,m$) to complex values, we obtain a function $\tilde{\rho}(\zeta_1,...,\zeta_m)$ defined on multiple complex variables $\zeta_1,...,\zeta_m$.
\section{Pseudoentropy sum rule}

\subsection{An approximated sum rule}
In \cite{Guo:2023aio} by using the formula (\ref{opsumold}) for pseudo-R\'enyi entropy, we derive a tentative sum rule for pseudoentropy, which is given by
\bea\label{sumpseudo}
S(\transA) = \sum_{k=0}^{2} \frac{\ep^{-\frac{2\pi \ii}{3}k }}{3\langle \phi |\psi\rangle} \frac{S(\rho_A(k))|_{n=1}}{[\mathcal{N}(\ep^{\frac{2\pi \ii}{3}k})]^{2}}.
\eea
However, it is shown in \cite{Guo:2023aio} the sum rule for pseudoentropy is not correct for general cases. It is only applicable if the distance between the two states $|\psi\rangle$ and $|\phi\rangle$ is small. For example $|\phi\rangle \propto |\psi\rangle+\epsilon |\psi'\rangle$, where $\epsilon\ll 1$, the above sum rule (\ref{sumpseudo}) is correct at the leading order of $\epsilon$.  This actually means the pseudoentropy also satisfy the first-law like relation at leading order of the perturbation \cite{Guo:2023aio}.

\subsection{A method to obtain entanglement entropy from R\'enyi entropy}
It is easy to check the sum rule (\ref{sumrulenew}) for $n$ being integers. But it would be subtle to analytically extend $n$ to complex numbers.  To avoid this, we would like to use a method to extract entanglement entropy from R\'enyi entropy without analytical continuation of $n$, which is introduced in \cite{DHoker:2020bcv}.

Firstly, let us briefly review this method. For a given density matrix $\rho$, suppose we know the trace of powers of $\rho$,
\bea
R_n(\rho)=tr \rho^n,
\eea
for the integers $n>1$. The strategy is to introduce the generating function 
\bea\label{generating}
G(z,\rho)=-tr \left(\rho\log\frac{1-z\rho}{1-z}\right)=\sum_{n=2}\frac{z^{n-1}}{n-1}\left[R_n(\rho)-1\right],
\eea
where the series is convergent in the unit disc $|z|<1$. The function $G(z,\rho)$ can be analytically continued to the cut plane $C\backslash [1,\infty)$, where $[1,\infty)$ is the branch cut of the logarithm. It can be shown one could obtain entanglement entropy by 
\bea
S(\rho):= -tr\rho\log \rho=\lim_{z\to -\infty}G(z,\rho).
\eea
For the application of this method, one may refer to the work \cite{DHoker:2020bcv}.

The above method can be generalized to the transition matrix $\mathcal{T}$, which is in general non-hermitian. One could also define the generating function $G(z,\mathcal{T})$ by replacing $\rho$ with $\mathcal{T}$ in (\ref{generating}). In our approach, the operator $\mathcal{T}$ would be normalized to $tr(\mathcal{T})=1$. One could similarly define the generating function 
\bea
G(z,\mathcal{T})=-tr \left(\mathcal{T}\log\frac{1-z\mathcal{T}}{1-z}\right).
\eea
Suppose the series expansion like (\ref{generating}) is convergent in the disc $|z|<1$. Then the function $G(z,\mathcal{T})$ can be analytically continued to the cut plane $C\backslash [1,\infty)$. Similarly, one could obtain the pseudoentropy by 
\bea
S(\mathcal{T}):=-tr \mathcal{T} \log \mathcal{T}=\lim_{z\to -\infty} G(z,\mathcal{T}).
\eea

\subsection{A tentative derivation of the sum rule}\label{sectionderivation}

By using the new sum rule (\ref{sumrulenew}) and the definition of generating function $G$, we have
\bea\label{generatingfunctionforPE}
&&G(z,\transA)=\sum_{n=2}^{\infty} \frac{z^{n-1}}{n-1}\left[\frac{1}{2\pi }\int_{-\pi}^\pi d\theta e^{-i n\theta}\mathcal{N}_\theta^n tr \rho_A(\theta)^n -1  \right]\nn \\
&&\phantom{G(z,\transA)}=\frac{1}{2\pi }\int_{-\pi}^\pi d\theta \sum_{n=2}^{\infty} \frac{z^{n-1}}{n-1} e^{-i n\theta} \mathcal{N}_\theta^n tr \rho_A(\theta)^n-\sum_{n=2}^{\infty} \frac{z^{n-1}}{n-1}\nn \\
&&\phantom{G(z,\transA)}=\frac{1}{2\pi }\int_{-\pi}^\pi d\theta G[z',\rho(\theta)]e^{-i\theta}\mathcal{N}_\theta+\frac{1}{2\pi}\int_{-\pi}^{\pi}d\theta \left[\sum_{n=2}^{\infty}\frac{z'^{n-1}}{n-1}e^{-i\theta}\mathcal{N}_\theta-\sum_{n=2}^{\infty} \frac{z^{n-1}}{n-1}\right],\nn \\
~
\eea
where $z':= z e^{-i\theta}\mathcal{N}_\theta$. Taking the limit $z\to -\infty$ on both side of the above equation, we have
\bea\label{integral1}
S(\transA)=\frac{1}{2\pi }\int_{-\pi}^\pi d\theta e^{-i\theta}\mathcal{N}_\theta S[\rho(\theta)] +\frac{1}{2\pi}\lim_{z\to-\infty}\int_{-\pi}^{\pi}d\theta \left[\sum_{n=2}^{\infty}\frac{z^{n-1}}{n-1}e^{-in\theta}\mathcal{N}_\theta^n-\sum_{n=2}^{\infty} \frac{z^{n-1}}{n-1}\right],
\eea
where we have assumed the order of integration and limit can be exchanged. Further we have
\bea
\int_{-\pi}^{\pi} d\theta e^{-i n \theta}\mathcal{N}_{\theta}^n=\int_{-\pi}^{\pi} d\theta e^{-in\theta}\left(2+e^{i\theta}\langle \phi |\psi\rangle +e^{-i\theta}\langle \psi| \phi\rangle\right)^n \langle \phi|\psi\rangle^{-n}=1.
\eea
Note that we have used the fact that $n$ is an integer. Thus the second term is vanishing.  Finally, we obtain the sum rule for pseudoentropy 
\bea\label{sumrulepseudo}
S(\transA)=\frac{1}{2\pi }\int_{-\pi}^\pi d\theta e^{-i\theta}\mathcal{N}_\theta S[\rho_A(\theta)].
\eea
As we will demonstrate below, the above sum rule for PE is only valid for some special cases.

The usual approach to obtain the EE or PE is through the analytical continuation of $n\to 1$.   
Actually, taking derivative with respect to $n$ on both side of (\ref{sumrulenew}) we would obtain a different result. By definition we have
$S(\transA)=-\partial_n tr(\transA)^n|_{n=1}$. On the right-hand side of (\ref{sumrulenew}) we would have
\bea
\frac{1}{2\pi} \int_{-\pi}^\pi d\theta e^{-i\theta} \mathcal{N}_{\theta}S[\rho_A(\theta)]+\frac{1}{2\pi} \int_{-\pi}^\pi d\theta e^{-i\theta}\mathcal{N}_\theta \log [e^{-i\theta}\mathcal{N}_\theta].
\eea
Comparing with (\ref{sumrulepseudo}), there exists an additional term that generally does not vanish. This implies that the integrand on the right-hand side of (\ref{sumrulenew}) cannot be analytically continued to arbitrary $n$. However, in the approach utilizing the generating functions $G(z,\mathcal{T})$ and $G(z,\rho(\theta))$, we consistently maintain $n$ as an integer.

\subsection{Methods by analytical continuation of superposition parameter}\label{sectionanalytical}
We can also express the sum rule by Cauchy integral. Define $\zeta=e^{i\theta}$, we have 
\bea\label{sumrulecomplex}
S(\transA)=\frac{1}{2\pi i} \oint_{|\zeta|= 1} \mathcal{N}_\zeta S(\rho_A(\zeta))d\zeta,
\eea
where $\mathcal{N}_\zeta:= \zeta^{-2}(2+\zeta\langle \phi |\psi\rangle +\zeta^{-1}\langle \psi| \phi\rangle )\langle \phi|\psi\rangle^{-1}$ and $S(\rho(\zeta)):=S(\rho_A(\theta))|_{\theta= -i\log \zeta}$. 
Here, we assume that the function $\mathcal{N}_\zeta S(\rho_A(\zeta))$ is a meromorphic function within the unit circle $|\zeta|\leq 1$. In general, $S(\rho_A(\zeta))$ may involve logarithmic functions, and it seems necessary to consider the branch cut of the logarithmic function for consistency. However, in certain cases, one can expand $S(\rho_A(\zeta))$ as polynomials of $\zeta$ with respect to certain parameters, such as the short interval length $\ell$ expansions in 2D CFT, as will be discussed in a later section. Thus, by the residue theorem, the PE can be expressed as
\bea
S(\transA)=\sum_{i,|\zeta|\le 1} \text{Res}\left[\mathcal{N}_\zeta S(\rho_A(\zeta)),a_i\right],
\eea
where Res$(f,a_i)$ denotes the residues of $f$ at $a_i$.

In Section \ref{superpositionsection}, we demonstrated that a more appropriate understanding of the operator sum rule (\ref{sumrulenew}) is achieved through the analytical continuation of the superposition parameter $\theta$. In fact, the form of the sum rule for PE (\ref{sumrulecomplex}) also implies a connection with the analytical continuation of $\theta$. 

\begin{figure}
\centering\includegraphics[width=5cm]{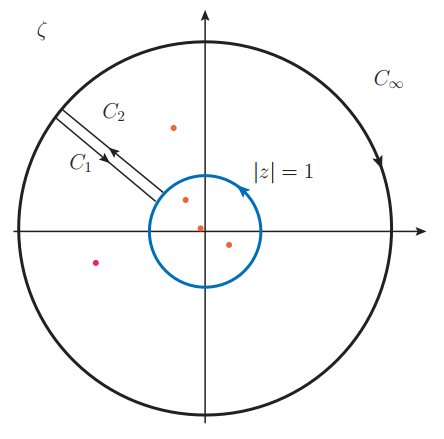}
\caption{The contours for Eq.(\ref{contourintegral}) and singularities (red dot) of the function $\mathcal{N}_\zeta S(\rho_A(\zeta))$.}\label{couterinfity}
\end{figure}

Recall the result of the operator $\tilde{\rho}(\zeta)$ defined on the complex $\zeta$ plane,
\bea
\rho(\zeta)=\mathcal{N}(\zeta)(|\phi\rangle+\zeta|\psi\rangle)(\langle\phi|+\frac{1}{\zeta}\langle\psi|),
\eea
where $\mathcal{N}(\zeta)=\zeta (2\zeta+\zeta^2 \langle \phi| \psi\rangle +\langle \psi|\phi\rangle)^{-1}$. As a function of $\zeta$, we find $\lim_{\zeta\to \infty}\rho(\zeta)=\trans$. 
The entropy function $S(\rho_A(\zeta))=-\text{tr}_A[\rho_A(\zeta)\log \rho_A(\zeta)]$ can also be considered as a function on the complex $\zeta$ plane. We find
\bea
\lim_{\zeta\to \infty}S(\rho_A(\zeta))=S(\transA).
\eea
This implies that the function $S(\rho_A(\zeta))$ approaches a constant as $\zeta$ tends to infinity. Eq.(\ref{sumrulecomplex}) suggests we can extract the PE by using Cauchy integral. Recall the function $\mathcal{N}_\zeta:= \zeta^{-2}(2+\zeta\langle \phi |\psi\rangle +\zeta^{-1}\langle \psi| \phi\rangle )\langle \phi|\psi\rangle^{-1}$ we find that at $\zeta\to\infty$ we have $\mathcal{N}_\zeta=\frac{1}{\zeta}+...$\;. We can choose a closed clockwise contour at infinity (as shown in Fig.\ref{couterinfity}) and extract the $S(\transA)$ by
\bea\label{PEatinfity}
S(\transA)=-\frac{1}{2\pi i}\oint_{C_\infty} d\zeta \mathcal{N}_\zeta S(\rho_A(\zeta)).
\eea
Note that the function $\mathcal{N}_\zeta S(\rho_A(\zeta))$ is expected to exhibit singularities on the $\zeta$ plane. As $\zeta  \to 0$, $\rho(\zeta)\to \frac{|\phi\rangle\langle \psi|}{\langle \psi|\psi\rangle}$, yet $\mathcal{N}_\zeta \to\frac{1}{\zeta^3}\frac{\langle \psi|\phi\rangle}{\langle \phi|\psi\rangle}$. Consequently, $\mathcal{N}_\zeta S(\rho_A(\zeta))$ is guaranteed to have at least one singularity at $\zeta=0$. The exact singularities of $S(\rho_A(\zeta))$ depends on states that we consider. In general, assume there exists singularities as shown in Fig.\ref{couterinfity}. By residue theorem we can express PE as the residue summations over the $\zeta$ plane,
\bea
S(\transA)=\sum_{i,\zeta\ \text{plane} } \text{Res}\left[\mathcal{N}_\zeta S(\rho_A(\zeta)),a_i\right].
\eea
This formula is valuable for evaluating the PE. However, our primary aim is to establish a connection between the PE $S(\transA)$ and the EE of $\rho_A(\theta)$. Therefore, one may use for the contours depicted in Fig.\ref{couterinfity}, as they facilitate this investigation. Again, by using residue theorem we have
\bea\label{contourintegral}
&&\frac{1}{2\pi i}\oint_{C_\infty} d\zeta \mathcal{N}_\zeta S(\rho_A(\zeta))+\frac{1}{2\pi i}\oint_{C_1} d\zeta \mathcal{N}_\zeta S(\rho_A(\zeta))+\frac{1}{2\pi i}\oint_{C_2} d\zeta \mathcal{N}_\zeta S(\rho_A(\zeta)) \nn \\
&&+\frac{1}{2\pi i}\oint_{|\zeta|=1} d\zeta \mathcal{N}_\zeta S(\rho_A(\zeta))=-\sum_{i,|\zeta|>1} \text{Res}\left[\mathcal{N}_\zeta S(\rho_A(\zeta)),a_i\right],
\eea
where $C_1$ and $C_2$ are two arbitrarily chosen paths that do not pass through any singularities. The contributions to the integral from $C_1$ and $C_2$ cancel each other out provided they are sufficiently close. By using (\ref{PEatinfity}) we find
\bea\label{PEsumrulefull}
S(\transA)=\frac{1}{2\pi i}\oint_{|\zeta|=1} d\zeta \mathcal{N}_\zeta S(\rho_A(\zeta))+\sum_{i,|\zeta|>1} \text{Res}\left[\mathcal{N}_\zeta S(\rho_A(\zeta)),a_i\right].
\eea
Comparing with formula (\ref{sumrulecomplex}), we observe additional contributions from the summation of residues in the region $|\zeta|>1$.
In the special case that there exists no singularities in $|\zeta|>1$. The formula (\ref{PEsumrulefull}) would reduce to (\ref{sumrulecomplex}). In the following sections, we will consider some examples that demonstrate the correctness of the sum rule formula (\ref{PEsumrulefull}) for general cases.

Before we move one let us comment the formula that we derived in section.\ref{sectionderivation}. In that approach there are no additional terms of the residues contributions (\ref{PEsumrulefull}). In the approach outlined in Section \ref{sectionderivation}, we make two assumptions. First, we assume that the order of integration and limit can be interchanged in Eq.(\ref{generatingfunctionforPE}). Second, we assume that the limit $\lim_{z\to -\infty}G[z',\rho(\theta)]$ with $z'=ze^{-i\theta}\mathcal{N}_\theta$ exists and equals $S(\rho_A(\theta))$. The function $G[z',\rho(\theta)]$ has a branch cut on $[1,\infty]$. When taking the limit $z\to -\infty$, we need to avoid the branch cut. However, $z'=ze^{-i\theta}\mathcal{N}_\theta$ may encounter the branch cut for certain values of $\theta$, which may pose a problem. Currently, we do not have a solution for this issue. Nonetheless, the derivation of (\ref{PEsumrulefull}) in this section remains valid.

\section{Examples}
To check the sum rule formula for PE we will consider some examples in this section.
\subsection{Qubit examples}\label{sectionqubit}
Assume the basis of two qubits system are $|00\rangle$, $|01\rangle$, $|10\rangle$ and $|11\rangle$. Define two pure states
\bea
&&|\phi\rangle= \frac{1}{\sqrt{ | \alpha_\phi|^2  + |\beta_\phi|^2}}(\alpha_\phi|00\rangle +\beta_\phi|11\rangle),\nn\\
&&|\psi\rangle= \frac{1}{\sqrt{ | \alpha_\psi|^2  + |\beta_\psi|^2}}(\alpha_\psi|00\rangle +\beta_\psi|11\rangle),
\eea
where $\alpha_\phi$, $\alpha_\psi$, $\beta_\phi$ and $\beta_\psi$ are arbitrary constants. The reduced density matrix of the superposition state $|\xi(\theta)\rangle=\mathcal{N}(\theta)(|\phi\rangle+e^{i\theta}|\psi\rangle) $ is diagonal, 
\begin{eqnarray}
\rho_A(\theta)=
\begin{pmatrix}
\lambda_1 & 0 \\
0 & \lambda_2
\end{pmatrix},
\end{eqnarray}
with
\bea
&& \lambda_1 = [\mathcal{N}(\theta)]^2
               \Big| \frac{\alpha _{\phi }}
                          {\sqrt{ |\alpha_{\phi}|^2 +|\beta_{\phi}|^2}}
                  + \frac{\alpha _{\psi } \ep^{i\theta}}
                         {\sqrt{ |\alpha_{\psi}|^2 +|\beta_{\psi}|^2}}
               \Big|^2, \nn\\
&& \lambda_2 = [\mathcal{N}(\theta)]^2
               \Big| \frac{\beta _{\phi }}
                          {\sqrt{ |\alpha_{\phi}|^2 +|\beta_{\phi}|^2}}
                    +\frac{\beta _{\psi } \ep^{i\theta}}
                          {\sqrt{ |\alpha_{\psi}|^2 +|\beta_{\psi}|^2}}\Big|^2, \nn
\eea
where
\bea
&& [\mathcal{N}(\theta)]^{-2}
  =\Big| \frac{\alpha _{\phi }}
                          {\sqrt{ |\alpha_{\phi}|^2 +|\beta_{\phi}|^2}}
                  + \frac{\alpha _{\psi } \ep^{i\theta}}
                         {\sqrt{ |\alpha_{\psi}|^2 +|\beta_{\psi}|^2}}
               \Big|^2 \nn\\
&& \phantom{[\mathcal{N}(\theta)]^{-2}
  =}+\Big| \frac{\beta _{\phi }}
                          {\sqrt{ |\alpha_{\phi}|^2 +|\beta_{\phi}|^2}}
                    +\frac{\beta _{\psi } \ep^{i\theta}}
                          {\sqrt{ |\alpha_{\psi}|^2 +|\beta_{\psi}|^2}}\Big|^2.
\eea
The reduced transition matrix $\transA$ is also diagonal, which is
\bea
\transA=\left(
\begin{array}{cc}
 t_1 & 0 \\
 0 & t_2 \\
\end{array}
\right),
\eea
with
\be
t_1=\frac{\a^*_\phi \alpha _{\psi }}{\a^*_\phi \alpha _{\psi }+\b^*_\phi \beta _{\psi }},
~~
t_2=\frac{\b^*_\phi \beta _{\psi }}{\a^*_\phi \alpha _{\psi }+\b^*_\phi \beta _{\psi }}.\nn
\ee
Here we consider $\alpha_\phi=\alpha_\psi=\beta_\phi=1$ and $\beta_\psi=e^{i\gamma}$. Assume $|\gamma|\ll 1$, thus we could expand the $S(\rho_A(\theta))$ as well as $S(\transA)$ with respect to parameter $\gamma$. With some calculation we have
\bea
S(\transA)=\log (2)+\frac{\gamma^2}{8}+\frac{\gamma^4}{64}+O\left(\gamma^5\right),
\eea
and
\bea
&&\mathcal{N}_\zeta S(\rho(\zeta))=\frac{(\zeta +1)^2 \log (2)}{\zeta ^3}-\frac{i \gamma  (\zeta +1) \log (2)}{\zeta ^3}+\frac{\gamma ^2 \left(\zeta ^2-2 \zeta +1-4 \log (2)\right)}{8 \zeta ^3}\nn \\
&&\phantom{\mathcal{N}_\zeta S(\rho(\zeta))} -\frac{i \gamma ^3 ((2\log (2)-3)\zeta  +3-4 \log (2))}{24 \zeta ^3}\nn \\
&&\phantom{\mathcal{N}_\zeta S(\rho(\zeta))}+\frac{\gamma ^4 \left(3 \zeta ^4+4 \zeta ^3+(10\log (2)-2)\zeta ^2 +4 (4\log (2)-5)\zeta  -9+10\log (2)\right)}{192 \zeta ^3 (\zeta +1)^2}+O(\gamma^5).\nn
\eea
One could evaluate the residues of the above formula in the unit circle,
\bea
\sum_{i,|\zeta|\le 1} \text{Res}\left[\mathcal{N}_\zeta S(\rho_A(\zeta)),a_i\right]=\log (2)+\frac{\gamma^2}{8}+\frac{\gamma^4}{64}+O\left(\gamma^5\right).
\eea
The sum rule formula (\ref{sumrulepseudo}) is correct for this case. One can compute the outcomes up to any desired order of $O(\gamma)$ and demonstrate the validity of the sum rule. Note that the sum rule for PE (\ref{sumpseudo}) derived in \cite{Guo:2023aio} is only valid up to the order of $O(\gamma^2)$.

\subsection{Perturbation state}\label{sectionperturbation}
Let us consider the two states $|\psi\rangle$ and $|\phi\rangle=\mathcal{N}_\phi(|\psi\rangle +\epsilon |\psi'\rangle)$, where $\mathcal{N}_\phi=1-\frac{1}{2}(\epsilon \langle \psi|\psi'\rangle +\epsilon^* \langle \psi'|\psi\rangle$+O($\epsilon^2$). We have
\bea
\langle \phi| \psi\rangle =1+\frac{1}{2}\epsilon^* \langle \psi'|\psi\rangle -\frac{1}{2}\epsilon \langle \psi |\psi'\rangle+O(\epsilon^2).
\eea
The reduced transition matrix is given by
\bea\label{perturbationT}
\transA=\rho_A^\psi-\epsilon^* \langle \psi'| \psi\rangle \rho_A^\psi+\epsilon^* \mathcal{T}_{A}^{\psi| \psi'}+O(\epsilon^2).
\eea 
The superposition state is
\bea
|\xi(\theta)\rangle= \mathcal{N}(\theta)\left[(1+e^{i\theta})|\psi\rangle-\frac{1}{2}(\epsilon \langle \psi|\psi' \rangle +\epsilon^* \langle \psi'| \psi\rangle)|\psi\rangle -\epsilon |\psi'\rangle \right]+O(\epsilon^2),
\eea
where the normalization satisfies
\bea
&&\mathcal{N}(\theta)^2|1+e^{i\theta}|^2=1+\frac{1}{2}\frac{1}{1+e^{i\theta}}(\epsilon \langle \psi|\psi' \rangle +\epsilon^* \langle \psi'| \psi\rangle)+\frac{1}{2}\frac{1}{1+e^{-i\theta}}(\epsilon \langle \psi|\psi' \rangle +\epsilon^* \langle \psi'| \psi\rangle)\nn \\
&&\phantom{\mathcal{N}(\theta)^2|1+e^{i\theta}|^2=}+\epsilon^* \frac{1}{1+e^{-i\theta}} \langle \psi'|\psi\rangle +\epsilon \frac{1}{1+e^{i\theta}}\langle \psi|\psi'\rangle+O(\epsilon^2).
\eea
The reduced density matrix $\rho_A(\theta)$ is given by 
\bea\label{perturbationtheta}
&&\rho_A(\theta)=\rho_A^\psi+\epsilon^* \frac{1}{1+e^{-i\theta}} \langle \psi'|\psi\rangle \rho_A^\psi-\epsilon^* \frac{1}{1+e^{-i\theta}} \mathcal{T}_A^{\psi|\psi'}\nn \\
&&\phantom{\rho_A(\theta)=\rho_A^\psi}+\epsilon \frac{1}{1+e^{i\theta}} \langle \psi|\psi'\rangle \rho_A^\psi-\epsilon \frac{1}{1+e^{i\theta}} \mathcal{T}_A^{\psi'|\psi}+O(\epsilon^2).
\eea
It would be straightforward to verify the correctness of the operator sum rule (\ref{sumrulenew}) at the leading order of $O(\epsilon)$.  In \cite{Guo:2023aio} we have shown actually the sum rule (\ref{sumpseudo}) is correct at the leading order of $O(\epsilon)$, but it can not be applicable at the order of $O(\epsilon^2)$.  One could evaluate the EE and PE at the leading order of $O(\epsilon)$ and check the sum rule for PE (\ref{PEsumrulefull}). 
Moreover, we would like to show it is also correct for the the order of $O(\epsilon^2)$.

\subsubsection{Perturbation calculation of EE and PE}
At the leading order of $O(\epsilon)$ both EE and PE satisfy the first-law like relation, we have
\bea\label{perturbationleadingT}
\delta S^{(1)}(\transA)= -\epsilon^* \langle \psi'| \psi\rangle \langle \psi| H_A^\psi| \psi\rangle +\epsilon^* \langle \psi'| H_A^\psi| \psi\rangle,
\eea 
and 
\bea\label{perturbationleadingrho}
&&\delta S^{(1)}(\rho(\theta))=\epsilon^* \frac{1}{1+e^{-i\theta}} \langle \psi'|\psi\rangle \langle \psi| H_A^\psi| \psi\rangle-\epsilon^* \frac{1}{1+e^{-i\theta}} \langle \psi'| H_A^\psi| \psi\rangle\nn \\
&&\phantom{\delta S^{(1)}(\rho(\theta))}+\epsilon \frac{1}{1+e^{i\theta}} \langle \psi|\psi'\rangle \langle \psi| H_A^\psi| \psi\rangle-\epsilon \frac{1}{1+e^{i\theta}} \langle \psi'| H_A^\psi| \psi\rangle,
\eea
where $H_A^\psi:=-\log \rho_A^\psi$ is the modular Hamiltonian for the state $\rho_A^\psi$. 

For an operator $M$ with $tr M=1$ we have the identity
\bea
-\log M=\int_0^\infty d\beta \left(\frac{1}{\beta+M}-\frac{1}{\beta+1} \right).
\eea
For our purposes, $M$ would stand for the reduced density matrix $\rho_A(\theta)$ (\ref{perturbationtheta}) or transition matrix $\transA$ (\ref{perturbationT}). We also have the   entropy formula
\bea
S(M)=-\tr M\log M=\int_0^\infty d\beta \left(tr M\frac{1}{\beta+M}-\frac{1}{\beta+1} \right).
\eea
We want to consider the second order variation of $S$ under the perturbation $M+\delta M$ with $tr \delta M=0$. The result is
\bea
\delta S^{(2)}(M)=-\int_0^\infty d\beta \beta \tr\left(\frac{1}{(\beta+M)^2}\delta M \frac{1}{\beta+M}\delta M \right),
\eea
see Appendix.\ref{appendixperturbation} for details.
Now the result can be used for the operator $\transA$ (\ref{perturbationT}) and $\rho(\theta)$ (\ref{perturbationtheta}). There is a subtle point. Typically, the spectra of the reduced transition matrix are complex. If $\transA$ exhibits negative spectra, the expression above might become undefined. Hence, it's preferable to assume that $\transA$ does not possess negative spectra.  We have
\bea\label{perturbationsecondT}
&&\delta S^{(2)}(\transA)
=-(\epsilon^*)^2\langle \psi'| \psi\rangle^2\int_0^\infty d\beta \beta \tr\left(\frac{1}{(\beta+\rho_A^\psi)^2}  \rho_A^\psi \frac{1}{\beta+\rho_A^\psi}  \rho_A^\psi \right)\nn \\
&&\phantom{\delta S^{(2)}(\transA)=}+(\epsilon^*)^2\langle \psi'| \psi\rangle\int_0^\infty d\beta \beta \tr\left(\frac{1}{(\beta+\rho_A^\psi)^2}  \rho_A^\psi \frac{1}{\beta+\rho_A^\psi}  \mathcal{T}^{\psi|\psi'} \right)\nn \\
&&\phantom{\delta S^{(2)}(\transA)=}+(\epsilon^*)^2\langle \psi'| \psi\rangle\int_0^\infty d\beta \beta \tr\left(\frac{1}{(\beta+\rho_A^\psi)^2}   \mathcal{T}^{\psi|\psi'} \frac{1}{\beta+\rho_A^\psi} \rho_A^\psi  \right)\nn \\
&&\phantom{\delta S^{(2)}(\transA)=}-(\epsilon^*)^2 \int_0^\infty d\beta \beta \tr\left(\frac{1}{(\beta+\rho_A^\psi)^2}  \mathcal{T}^{\psi|\psi'} \frac{1}{\beta+\rho_A^\psi} \mathcal{T}^{\psi|\psi'} \right),
\eea
and
\bea\label{perturbationsecondtheta}
&&\delta S^{(2)}(\rho_A(\theta))=-(\epsilon^*)^2\frac{\langle \psi'| \psi\rangle^2}{(1+e^{-i\theta})^2}\int_0^\infty d\beta \beta \tr\left(\frac{1}{(\beta+\rho_A^\psi)^2}  \rho_A^\psi \frac{1}{\beta+\rho_A^\psi}  \rho_A^\psi \right)\nn \\
&&\phantom{\delta S^{(2)}(\rho_A(\theta))=}+(\epsilon^*)^2\frac{\langle \psi'| \psi\rangle}{(1+e^{-i\theta})^2}\int_0^\infty d\beta \beta \tr\left(\frac{1}{(\beta+\rho_A^\psi)^2}  \rho_A^\psi \frac{1}{\beta+\rho_A^\psi}  \mathcal{T}^{\psi|\psi'} \right)\nn \\
&&\phantom{\delta S^{(2)}(\rho_A(\theta))=}+(\epsilon^*)^2\frac{\langle \psi'| \psi\rangle}{(1+e^{-i\theta})^2}\int_0^\infty d\beta \beta \tr\left(\frac{1}{(\beta+\rho_A^\psi)^2}   \mathcal{T}^{\psi|\psi'} \frac{1}{\beta+\rho_A^\psi} \rho_A^\psi  \right)\nn \\
&&\phantom{\delta S^{(2)}(\rho_A(\theta))=}-(\epsilon^*)^2 \frac{1}{(1+e^{-i\theta})^2}\int_0^\infty d\beta \beta \tr\left(\frac{1}{(\beta+\rho_A^\psi)^2}  \mathcal{T}^{\psi|\psi'} \frac{1}{\beta+\rho_A^\psi} \mathcal{T}^{\psi|\psi'} \right)\nn \\
&&\phantom{\delta S^{(2)}(\rho_A(\theta))=}+h.c.,
\eea
\subsubsection{Sum rule for perturbation state}
Now using the results (\ref{perturbationsecondT}) (\ref{perturbationsecondtheta}), one could check the sum rule (\ref{sumrulepseudo}) or (\ref{sumrulecomplex}). By definition we have
\bea\label{normperturbation}
&&\mathcal{N}_\theta=(2+e^{i\theta}\langle \phi|\psi\rangle+e^{-i\theta}\langle\psi| \phi\rangle)\langle \phi|\psi\rangle^{-1}\nn \\
&&\phantom{\mathcal{N}_\theta}=2+e^{i\theta}+e^{-i\theta}-\epsilon^* \langle \psi'|\psi\rangle (1+e^{-i\theta})+\epsilon \langle \psi|\psi'\rangle (1+e^{-i\theta})+O(\epsilon^2).\nn
\eea
At the leading order of $O(\epsilon)$  by using (\ref{perturbationleadingT}) (\ref{perturbationleadingrho}) and (\ref{normperturbation}) we have
\bea
&&\frac{1}{2\pi}\int_{-\pi}^\pi d\theta e^{-i\theta}(2+e^{i\theta}+e^{-i\theta})\delta S^{(1)}(\rho(\theta))\nn \\
&&=\frac{1}{2\pi}\int_{-\pi}^\pi d\theta (1+e^{-i\theta})^2\delta S^{(1)}(\rho(\theta))=\delta S^{(1)}(\transA).
\eea
This result demonstrates the validity of the sum rule (\ref{sumrulepseudo}) at the leading order of $O(\epsilon)$. To investigate the order of $O(\epsilon^2)$, we must assess two terms. The first involves the multiplication of the $O(\epsilon)$ component of $\mathcal{N}_\theta$ (\ref{normperturbation}) with $\delta S^{(1)}$, that is
\bea
\frac{1}{2\pi}\int_{-\pi}^\pi d\theta e^{-i\theta}(-\epsilon^* \langle \psi'|\psi\rangle (1+e^{-i\theta})+\epsilon \langle \psi|\psi'\rangle (1+e^{-i\theta}))\delta S^{(1)}(\rho(\theta)),
\eea
which is vanishing. The other term is
\bea
\frac{1}{2\pi}\int_{-\pi}^\pi d\theta e^{-i\theta}(2+e^{i\theta}+e^{-i\theta})\delta S^{(2)}(\rho(\theta))=\delta S^{(2)}(\transA),
\eea
which can be confirmed through straightforward calculations. Thus, we have established the validity of the sum rule for the PE (\ref{sumrulepseudo}) for the perturbed state up to the order of $O(\epsilon^2)$. If we analytically continue $\theta$ to the complex plane $\zeta$, for the perturbed state, there are no singularities in the region where $|\zeta|>1$. Consequently, the sum rule given by equation (\ref{sumrulepseudo}) is equivalent to the sum rule given by equation (\ref{PEsumrulefull}).  Furthermore, one could extend these calculations to demonstrate its correctness for any order of perturbation.

\subsection{Short interval expansion}
Consider an interval $A$ with length $\ell$ in 2-dimensional CFT . One could evaluate the pseudoentropy and EE by using operator product expansion (OPE) of twist operator \cite{Cardy:2007mb}-\cite{Chen:2013kpa}. For the reduced density matrix $\rho_A$ we have the expansion
\bea
S_A(\rho_A)=\frac{c}{3}\log\frac{\ell}{\epsilon}+\sum_{m=1}^n \sum_{\mathcal{X}_1,...,\mX_m} \ell^{\Delta_{\mathcal{X}_1}+...+\Delta_{\mX_m}} a_{\mathcal{X}_1...\mX_m} tr(\rho_A \mathcal{X}_1)...tr(\rho_A \mX_m),
\eea
where $\mX_i$ ($i=1,...,m$) are the quasi-primary operators for the OPE of twist operator, $\Delta_{\mX_i}$ are their conformal dimension,  $h_\sigma$ is the conformal weight of twist operator, $a_{\m\mathcal{X}_1...\mX_m}$ are the constant coefficients that are independent with $\rho_A$. For the transition matrix $\transA$ we also have the similar expansion 
\bea
S_A(\transA)=\frac{c}{3}\log\frac{\ell}{\epsilon}+\sum_{m=1}^n \sum_{\mathcal{X}_1,...,\mX_m} \ell^{\Delta_{\mathcal{X}_1}+...+\Delta_{\mX_m}} a_{\mathcal{X}_1...\mX_m} tr(\transA \mathcal{X}_1)...tr(\transA \mX_m).
\eea
Define the function $F_{m}(\theta):=tr(\rho_A(\theta) \m\mathcal{X}_1)...tr(\rho(\theta)_A \mX_m)$, which can be expanded as
\bea
F_{m}(\theta)=\mathcal{N}(\theta)^{2m}\prod_{i=1}^m \left(\langle \phi| \mX_i |\phi\rangle+ e^{-i\theta}\langle \psi|\mX_i|\phi\rangle+e^{i\theta}\langle \phi| \mX_i|\psi\rangle +\langle \psi| \mX_i |\psi\rangle\right),
\eea
Note that the coefficients $a_{\m\mathcal{X}_1...\mX_m}$ are same for the PE and EE. To check the sum rule we only need to consider the following integral,
\bea\label{integration}
&&\frac{1}{2\pi}\int_{-\pi}^{\pi}d\theta e^{-i\theta}\mathcal{N}_\theta F_m(\theta)\nn \\
&&=\frac{1}{2\pi}\int_{-\pi}^{\pi}d\theta e^{-i\theta}
\frac{\prod_{i=1}^m \left(\langle \phi| \mX_i |\phi\rangle+ e^{-i\theta}\langle \psi|\mX_i|\phi\rangle+e^{i\theta}\langle \phi| \mX_i|\psi\rangle +\langle \psi| \mX_i |\psi\rangle\right)}{\langle \phi|\psi\rangle \left(2+e^{i\theta}\langle \phi|\psi\rangle+e^{-i\theta}\langle\psi| \phi\rangle \right)^{m-1}}\nn
\eea
With $\zeta=e^{i\theta}$ we have
\bea
\frac{1}{2\pi}\int_{-\pi}^{\pi}d\theta e^{-i\theta}\mathcal{N}_\theta F_m(\theta)= \frac{1}{2\pi i} \oint_{|\zeta|=1} d\zeta \mathcal{F}_m(\zeta),\nn
\eea
with 
\bea
\mathcal{F}_m(\zeta):=\frac{1}{\zeta^3}\frac{\prod_{i=1}^m \left(\zeta\langle \phi| \mX_i |\phi\rangle+ \langle \psi|\mX_i|\phi\rangle+\zeta^2\langle \phi| \mX_i|\psi\rangle +\zeta\langle \psi| \mX_i |\psi\rangle\right)}{\langle \phi|\psi\rangle \left(2\zeta+\zeta^2\langle \phi|\psi\rangle+\langle\psi| \phi\rangle \right)^{m-1}}.
\eea
To compute the integration, we must identify the singularities of the integrand within the unit circle $|\zeta| \leq 1$. Utilizing the Cauchy–Schwarz inequality, we have $|\langle \phi|\psi\rangle|\le 1$. For $m\ge 1$, within the unit circle, there are two singularities at $\zeta=0$ and $\zeta=\frac{-1+\sqrt{1-|\langle \phi|\psi\rangle|^2}}{\langle \phi|\psi\rangle}$. Note that the integrand also has a singularity at $\zeta=\frac{-1-\sqrt{1-|\langle \phi|\psi\rangle|^2}}{\langle \phi|\psi\rangle}$.
\subsection{Results for $m=1,2$}
Now, we aim to evaluate the integral (\ref{integration}). Firstly, let's consider $m=1$. In this case, $\zeta=0$ is the only singularity. The result is
\bea
\frac{1}{2\pi}\int_{-\pi}^{\pi}d\theta e^{-i\theta}\mathcal{N}_\theta F_1(\theta)=\frac{\langle \phi |\mathcal{X}_1|\psi\rangle}{\langle \phi|\psi\rangle},
\eea
which is consistent with the sum rule (\ref{sumrulepseudo}). Then, let us consider $m=2$. As noted above there are two singularities $\zeta_0=0$ and $\zeta_1=\frac{-1+\sqrt{1-|\langle \phi|\psi\rangle|^2}}{\langle \phi|\psi\rangle}$ within the unit circle $|\zeta|\le 1$, while the singularity $\zeta_2=\frac{-1-\sqrt{1-|\langle \phi|\psi\rangle|^2}}{\langle \phi|\psi\rangle}$ is in the region $|\zeta|\ge 1$. Thus we have
\bea\label{integrationF2}
&&\frac{1}{2\pi}\int_{-\pi}^{\pi}d\theta e^{-i\theta}\mathcal{N}_\theta F_2(\theta)=\sum_{j=0,1}\text{Res}[\mathcal{F}_2(\zeta),\zeta_j]\nn 
\eea
With some calculations we find
\bea
&&\frac{1}{2\pi}\int_{-\pi}^{\pi}d\theta e^{-i\theta}\mathcal{N}_\theta F_2(\theta)=-\frac{1}{\zeta _1^3 \left(\zeta _1-\zeta _2\right) \langle \phi |\psi \rangle }\Big[\left\langle \psi \left|\mathcal{X}_1\right|\phi \right\rangle  \left\langle \psi \left|\mathcal{X}_2\right|\phi \right\rangle\nn \\
&&+\zeta _1 \left(\left\langle \phi \left|\mathcal{X}_1\right|\phi \right\rangle  \left\langle \psi \left|\mathcal{X}_2\right|\phi \right\rangle +\left\langle \psi \left|\mathcal{X}_1\right|\psi \right\rangle  \left\langle \psi \left|\mathcal{X}_2\right|\phi \right\rangle +\left\langle \psi \left|\mathcal{X}_1\right|\phi \right\rangle  \left(\left\langle \psi \left|\mathcal{X}_2\right|\psi \right\rangle +\left\langle \phi \left|\mathcal{X}_2\right|\phi \right\rangle \right)\right)\nn \\
&&\zeta _1^2 \big(\left\langle \phi \left|\mathcal{X}_2\right|\psi \right\rangle  \left\langle \psi \left|\mathcal{X}_1\right|\phi \right\rangle +\left\langle \phi \left|\mathcal{X}_1\right|\psi \right\rangle  \left\langle \psi \left|\mathcal{X}_2\right|\phi \right\rangle +\left\langle \phi \left|\mathcal{X}_1\right|\phi \right\rangle  \left(\left\langle \psi \left|\mathcal{X}_2\right|\psi \right\rangle +\left\langle \phi \left|\mathcal{X}_2\right|\phi \right\rangle \right)\nn \\
&&+\left\langle \psi \left|\mathcal{X}_1\right|\psi \right\rangle  \left(\left\langle \psi \left|\mathcal{X}_2\right|\psi \right\rangle +\left\langle \phi \left|\mathcal{X}_2\right|\phi \right\rangle \right)\big)+\zeta _1^3 \big((\left\langle \phi \left|\mathcal{X}_2\right|\psi \right\rangle  \left(\left\langle \phi \left|\mathcal{X}_1\right|\phi \right\rangle +\left\langle \psi \left|\mathcal{X}_1\right|\psi \right\rangle \right)\nn \\
&&+\left\langle \phi \left|\mathcal{X}_1\right|\psi \right\rangle  \left(\left\langle \phi \left|\mathcal{X}_2\right|\phi \right\rangle +\left\langle \psi \left|\mathcal{X}_2\right|\psi \right\rangle \right))+\left\langle \phi \left|\mathcal{X}_1\right|\psi \right\rangle  \left\langle \phi \left|\mathcal{X}_2\right|\psi \right\rangle  \zeta _2\big)\Big].
\eea
This result indicates that the sum rule (\ref{sumrulepseudo}) is not valid in this instance. However, formula (\ref{PEsumrulefull}) stands as the correct one. We can calculate the contributions of the singularity at point $\zeta_2$, and it is straightforward to verify that
\bea
\frac{1}{2\pi}\int_{-\pi}^{\pi}d\theta e^{-i\theta}\mathcal{N}_\theta F_2(\theta)+\text{Res}[\mathcal{F}_2(\zeta),\zeta_2]=\frac{\langle\phi |\mathcal{X}_1|\psi\rangle \langle\phi |\mathcal{X}_2|\psi\rangle}{\langle \phi|\psi\rangle^2},
\eea
the right-hand of which yields the anticipated outcome for the short interval expansion of PE. 
\subsection{Results for arbitrary $m$}\label{sectioncorrelatorsumrulealt}
Indeed, one could verify the outcomes for $m>2$ through direct calculations. Here we show that one could obtain the results for any $m$ by the property of the function $\mathcal{F}_m(\zeta)$. The argument here is analogous to the general proof presented in section.\ref{sectionanalytical}. In the limit $\zeta\to \infty$ we find
\bea
\lim_{\zeta\to \infty}\mathcal{F}_m(\zeta) \to \frac{1}{\zeta}\frac{\prod_{i=1}^m \langle \phi|\mX_i|\psi\rangle}{\langle \phi|\psi\rangle^m}.
\eea
Note that the coefficient of $\frac{1}{\zeta}$ in the above formula precisely represents the term for evaluating PE. Therefore, this term can be isolated by integrating along the circle $C_\infty$ at infinity, as illustrated in Fig. \ref{couterinfity}. For all functions $\mathcal{F}m(\zeta)$, there exist three singularities ${\zeta_0,\zeta_1,\zeta_2 }$, as discussed in previous sections. Hence, it is feasible to rewrite the integral along the circle $C_\infty$ as a summation of residues on the $\zeta$ plane.
\bea
\frac{\prod_{i=1}^m \langle \phi|\mX_i|\psi\rangle}{\langle \phi|\psi\rangle^m}=\sum_{i,\zeta\ \text{plane} } \text{Res}\left[\mathcal{F}_m,\zeta_i\right].
\eea
While the residues summations for $\zeta_0$ and $\zeta_1$ can be rewritten as integral on the unit circle $|\zeta|=1$. As a result we have
\bea\label{sumrulecorrelatoralt}
\frac{1}{2\pi}\int_{-\pi}^{\pi}d\theta e^{-i\theta}\mathcal{N}_\theta F_m(\theta)+\text{Res}[\mathcal{F}_m(\zeta),\zeta_2]=\frac{\prod_{i=1}^m \langle \phi|\mX_i|\psi\rangle}{\langle \phi|\psi\rangle^m}.
\eea
By employing the above formula, one can demonstrate the sum rule for PE (\ref{PEsumrulefull}) in the short interval expansion. 

\section{Applications}\label{sectionapplication}

\subsection{Holographic dual of transition matrix}
The transition matrix is typically non-Hermitian, resulting in its spectra, as well as pseudo-Rényi entropy and PE, being complex. It is anticipated that there exists a subset of transition matrices whose spectra are positive. For these states, both pseudo-Rényi entropy and PE are positive. In the context of AdS/CFT, this subset of transition matrices may have a bulk dual. Given that the calculation method for pseudo-Rényi entropy and PE in QFTs closely parallels that of Rényi entropy and EE, it is reasonable to anticipate that if a given transition matrix $\trans$ possesses a bulk geometric dual, then pseudo-Rényi entropy and PE may also have corresponding bulk geometric interpretations. For PE, its bulk dual is similarly described by the RT formula.

In \cite{Guo:2022jzs}, the authors utilize the concept of pseudo-Hermiticity to classify transition matrices. They particularly focus on a class of transition matrices $\trans$ for which $|\phi\rangle = \eta |\psi\rangle$, with $\eta$ being a Hermitian and invertible operator. In such cases where the transition matrix exhibits pseudo-Hermitian properties, it possesses certain distinctive characteristics. Moreover, if $\eta$ can be expressed as an integration or summation of local operators, then for any subsystem $A$, $\eta$ can be decomposed into $\eta_A\otimes\eta_{\bar A}$.

It is further demonstrated in \cite{Guo:2022jzs} that if both $\eta_A$ and $\eta_{\bar A}$ are either positive or negative operators, the reduced transition matrix $\transA$ should exhibit positive spectra. Consequently, the pseudo-Rényi entropy and PE also yield positive values. Transition matrices of this kind may potentially possess a bulk geometric dual.

The sum rules that we explored in this paper may have interesting geometric explanation if the transition matrix can be dual to a bulk geometry.
In the traditional treatment the bulk geometry is considered to be associated with boundary state with a given density matrix.   It is generally expected the theory can be dual to gravity should be a gapped large-N QFTs\cite{Aharony:1999ti}\cite{Harlow:2018fse} with bulk Newton constant $G\sim 1/N^2$. 
The semi-classical limit $G\to 0$ corresponds to the large $N$ limit in the dual QFT. If a state $|\Psi\rangle$ has a bulk geometric dual\footnote{Generally, it is not anticipated that a pure single state can possess a bulk geometric dual. Instead, it may comprise a superposition of numerous pure states, all sharing the same property as $|\Psi\rangle$ in the semi-classical limit. For simplicity, we employ the notation of a single pure state $|\Psi\rangle$ to represent the states dual to a specified bulk geometry. }, the connected 2-point correlation functions of any single trace operator $\mathcal{O}_i$ should satisfy the following relation
\bea\label{connected_two}
\langle \Psi| \mathcal{O}_i \mathcal{O}_j |\Psi\rangle-\langle \Psi| \mathcal{O}_i|\Psi\rangle \langle\Psi| \mathcal{O}_j |\Psi\rangle\sim O(N^2).
\eea 
Let us assume the state $\rho(\theta):=|\xi(\theta)\rangle \langle \xi(\theta)|$ can be dual to bulk geometry. Thus the single trace operators satisfy the relation (\ref{connected_two}) for $|\xi(\theta)\rangle$. Now using the sum rule for the correlators (\ref{correlatorsumrule}), we have
\bea
&&\frac{\langle \phi| \mO_i \mO_j|\psi\rangle -\langle \phi| \mO_i |\psi\rangle \langle \phi| \mO_j|\psi\rangle}{\langle \phi|\psi\rangle^2}\nn \\
&&=\frac{1}{2\pi}\int_{-\pi}^\pi d\theta e^{-i n\theta}\mathcal{N}_\theta^{n} \left(\langle \xi(\theta)| \mathcal{O}_i \mathcal{O}_j |\xi(\theta)\rangle-\langle \xi(\theta)| \mathcal{O}_i|\xi(\theta)\rangle \langle\xi(\theta)| \mathcal{O}_j |\xi(\theta)\rangle\right).\nn
\eea
Notice that the term inside the brackets scales as $O(N^2)$. Consequently, the left-hand side of the equation above should also scale as $O(N^2)$. This reasoning extends to $n$-point correlation functions. The sum rule (\ref{correlatorsumrule}) ensures that both the superposition state $\rho(\theta)$ and the transition matrix $\trans$ exhibit the same scaling behavior in the large $N$ limit. Therefore, if the state $\rho(\theta)$ can be effectively described by a bulk geometry, it suggests the possibility that the transition matrix $\trans$ is also dual to some bulk geometry.

In fact, the sum rule $(\ref{correlatorsumrule})$ provides a link between two geometries. The bulk metric $g_{\mu\nu}$ is typically correlated with the boundary expectation value of the stress-energy tensor. By applying the sum rule $(\ref{correlatorsumrule})$ to the stress-energy tensor, one can establish a relationship between the two bulk metrics.

Moreover, the sum rule $(\ref{pseudoRenyisumrule})$ can establish a connection between the bulk on-shell action. In line with the holographic dual proposal for R\'enyi entropy, the evaluation of R\'enyi entropy can be interpreted as the assessment of bulk on-shell action, incorporating a cosmic brane with tension $T_n=\frac{n-1}{4n G}$ insertion\cite{Dong:2016fnf}. This proposal extends to the transition matrix scenario. In the semiclassical limit $G\to 0$, we attain the relationship:
\bea
tr(\transA)^n=e^{-(1-n)S^{(n)}(\transA)}\simeq e^{-I_n(\transA)},
\eea 
Here, $I_n$ represents the on-shell action of the bulk metric considering the backreaction of the cosmic brane. On the right-hand side of (\ref{pseudoRenyisumrule}), employing a similar argument, we obtain
\bea
\frac{1}{2\pi}\int_{-\pi}^\pi d\theta e^{-i n\theta+n \log \mathcal{N}_\theta -I_n(\rho_A(\theta))}.
\eea
The above integration can be assessed utilizing the saddle point approximation, which involves solving the equation
\bea\label{saddle}
\partial_\theta [-i n\theta+n \log \mathcal{N}_\theta -I_n(\rho_A(\theta))]=0.
\eea
Assume the above equation has one solution $\theta^*$. The sum rule now can be written as the relation between two on-shell action
\bea
e^{-I_n(\transA)}\simeq e^{-i n\theta^*+n \log \mathcal{N}_{\theta^*} -I_n(\rho_A(\theta^*))}.
\eea
If there are multiple solutions to (\ref{saddle}), all contributions from these solutions should be included.

\subsection{Bound of $|S(\transA)|$}

It is an intriguing question whether the real and imaginary parts of $S(\transA)$ possess lower or upper bounds. In \cite{Mollabashi:2020yie}, the author delves into the PE within free field theory. To motivate their investigation, they introduce the quantity (expressed using our notation) 
\bea\label{difference}
\Delta S_{\psi,\phi}:=|S(\transA)|-\frac{S(\rho_A^\psi)+S(\rho_A^\phi)}{2}.
\eea
It has been observed that this quantity is consistently non-positive in the examples they have considered. There is a conjecture that this property may be universal in QFTs, although there exist examples in qubit systems where this quantity is positive.

Our summation rule for PE (\ref{PEsumrulefull}) can be employed to determine the upper bound of PE. It's important to emphasize that this bound is universal, as the summation rule (\ref{PEsumrulefull}) holds true for arbitrary states $|\psi\rangle$ and $|\phi\rangle$.
By examining the expression (\ref{PEsumrulefull}), we observe that the bound of PE can be linked to the EE of the superposition state $|\xi(\theta)\rangle$. Indeed, the bound of a superposition state has been investigated in \cite{superposition}. We will adopt the approach outlined in that paper.

\subsubsection{A bound of $|S(\transA)|$}
We will use the following inequalities that hold for any density matrices $\rho_1$ and $\rho_2$ and $|\alpha_1|^2+|\alpha_2|^2=1$,
\bea\label{eq1}
|\alpha_1 |^2 S(\rho_1)+|\alpha_1 |^2 S(\rho_2)\le S(|\alpha_1|^2\rho_1+|\alpha_2|^2\rho_2)
\eea
and 
\bea\label{eq2}
S(|\alpha_1|^2\rho_1+|\alpha_2|^2\rho_2)\le |\alpha_1 |^2 S(\rho_1)+|\alpha_1 |^2 S(\rho_2)+ H(|\alpha_1|^2),
\eea
where $H(|\alpha_1|^2)=-|\alpha_1|^2\log |\alpha_1|^2-|\alpha_2|^2\log |\alpha_2|^2$.

Let us define the new superposition state
\bea\label{xiprime}
|\xi'(\theta)\rangle:= \mathcal{N}'(\theta)(|\phi\rangle-e^{i\theta}|\psi\rangle),
\eea
where $\mathcal{N}'(\theta):=1/\sqrt{2-e^{i\theta} \langle \phi| \psi\rangle -e^{-i\theta} \langle \psi| \phi\rangle}$. With some calculations we have
\bea\label{density}
\frac{1}{2}\rho_A(\phi)+\frac{1}{2}\rho_A(\psi)=\frac{1}{4 \mathcal{N}(\theta)^2}\rho_A(\theta)+\frac{1}{4 \mathcal{N}'(\theta)^2}\rho'_A(\theta),
\eea
where the reduced density matrices are defined as
$\rho_A(\psi):= tr_{\bar A}|\psi\rangle \langle \psi|$, $\rho_A(\phi):=tr_{\bar A}|\phi\rangle \langle \phi|$ and $\rho'_A(\theta):=tr_{\bar A}|\xi'(\theta)\rangle \langle \xi'(\theta)|$.
Using (\ref{eq2}) for the left hand side of (\ref{density}), we have
\bea
S(\frac{1}{2}\rho_A(\phi)+\frac{1}{2}\rho_A(\psi))\le \frac{1}{2}S(\rho_A(\phi))+\frac{1}{2}S(\rho_A(\psi))+\log 2.
\eea
While using (\ref{eq1}) for the right hand side of (\ref{density}) we get
\bea
\frac{1}{4 \mathcal{N}(\theta)^2}S(\rho_A(\theta))+\frac{1}{4 \mathcal{N}'(\theta)^2} S(\rho'_A(\theta))\le S\left( \frac{1}{4 \mathcal{N}(\theta)^2}\rho_A(\theta)+\frac{1}{4 \mathcal{N}'(\theta)^2}\rho'_A(\theta)\right).
\eea
Therefore, we find
\bea\label{inequa}
\frac{1}{2 \mathcal{N}(\theta)^2}S(\rho_A(\theta))+\frac{1}{2 \mathcal{N}'(\theta)^2} S(\rho'_A(\theta))\le  S(\rho_A(\phi))+S(\rho_A(\psi))+2\log 2.
\eea
The sum rule (\ref{PEsumrulefull}) can be written as 
\bea
S(\transA)=\frac{1}{2\pi }\int_{-\pi}^\pi d\theta e^{-i\theta}\mathcal{N}_\theta S[\rho_A(\theta)]+\sum_{i,|\zeta|>1} \text{Res}\left[\mathcal{N}_\zeta S(\rho_A(\zeta)),a_i\right].
\eea
We can obtain an upper bound of $|S(\transA)|$,
\bea\label{upperbound}
&&|S(\transA)|\le \frac{1}{2\pi}\int_{-\pi}^\pi d\theta \mathcal{N}(\theta)^{-2}|\langle \phi| \psi\rangle|^{-1}S(\rho_A(\theta))+|\sum_{i,|\zeta|>1} \text{Res}\left[\mathcal{N}_\zeta S(\rho_A(\zeta)),a_i\right]|\nn \\
&&\phantom{|S(\transA)|}=\frac{1}{2\pi}\int_{-\pi}^0 d\theta \mathcal{N}(\theta)^{-2}|\langle \phi| \psi\rangle|^{-1}S(\rho_A(\theta))+\frac{1}{2\pi}\int_{-\pi}^0 d\theta \mathcal{N}'(\theta)^{-2}|\langle \phi| \psi\rangle|^{-1}S(\rho'_A(\theta))\nn\\
&&\phantom{|S(\transA)|=}+|\sum_{i,|\zeta|>1} \text{Res}\left[\mathcal{N}_\zeta S(\rho_A(\zeta)),a_i\right]|\nn \\
&&\phantom{|S(\transA)|}\le \frac{  S(\rho_A(\phi))+S(\rho_A(\psi))+2\log 2}{|\langle \phi|\psi\rangle|}+|\sum_{i,|\zeta|>1} \text{Res}\left[\mathcal{N}_\zeta S(\rho_A(\zeta)),a_i\right]|,
\eea
where we use  the fact $\mathcal{N}(\theta+\pi)=\mathcal{N}'(\theta)$, $S(\rho_A(\theta+\pi))=S(\rho'_A(\theta))$ in the second step and (\ref{inequa}) in the third step.

The expression (\ref{upperbound}) is useless for general case since it involves a summation of residues, which can only be determined once the function $\mathcal{N}_\zeta S(\rho_A(\zeta))$ is known. In the previous section, we demonstrated that there are cases where the function $\mathcal{N}_\zeta S(\rho_A(\zeta))$ exhibits no singularities within the region $|\zeta|>1$. In such instances, the absolute value of $S(\transA)$ is bounded by the quantities $S(\rho_A^\psi)$, $S(\rho_A^\phi)$, and $|\langle \phi|\psi\rangle|$.

\subsubsection{Example}
In section.\ref{sectionperturbation} we discuss the PE and sum rule for perturbed states. We establish that there are no singularities in the region $|\zeta| > 1$ up to the order of $O(\epsilon^2)$. Thus we don't need to consider the term of the residues summation of the upper bound (\ref{upperbound}). In this case it follows that $|\langle \phi|\psi\rangle|=1+O(\epsilon)$. We also find $S(\rho_A^\phi)=S(\rho_A^\psi)+O(\epsilon)$ and $S(\transA)=S(\rho_A^\psi)+O(\epsilon)$. It is evident that the inequality (\ref{upperbound}) is satisfied in this case.

Let us consider the two qubits system discussed in section.\ref{sectionqubit} as another example. We define the function
\bea
\Delta \tilde{S}_{\psi,\phi}:=|S(\transA)|-\frac{  S(\rho_A(\phi))+S(\rho_A(\psi))+2\log 2}{|\langle \phi|\psi\rangle|}.
\eea
\begin{figure}[htbp]
  \centering
  \begin{minipage}[b]{0.45\textwidth} 
    \includegraphics[width=\textwidth]{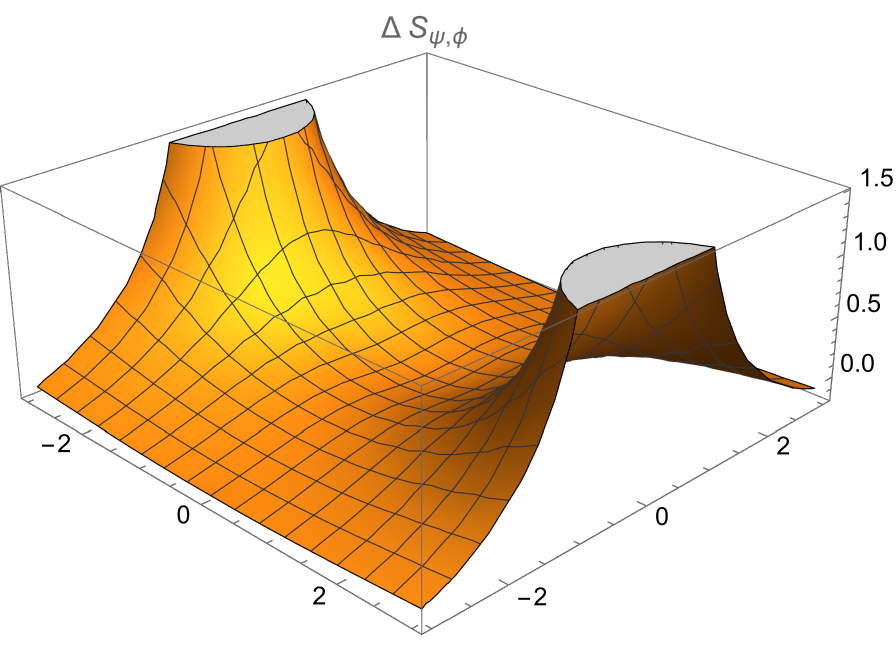}
    \caption{Plot of  $\Delta S_{\psi,\phi}$ as a function of the parameter $\gamma$, where both the real and imaginary parts of $\gamma$ range from $-3$ to $3$. }\label{F1}
  \end{minipage}
  \hfill 
  \begin{minipage}[b]{0.45\textwidth} 
    \includegraphics[width=\textwidth]{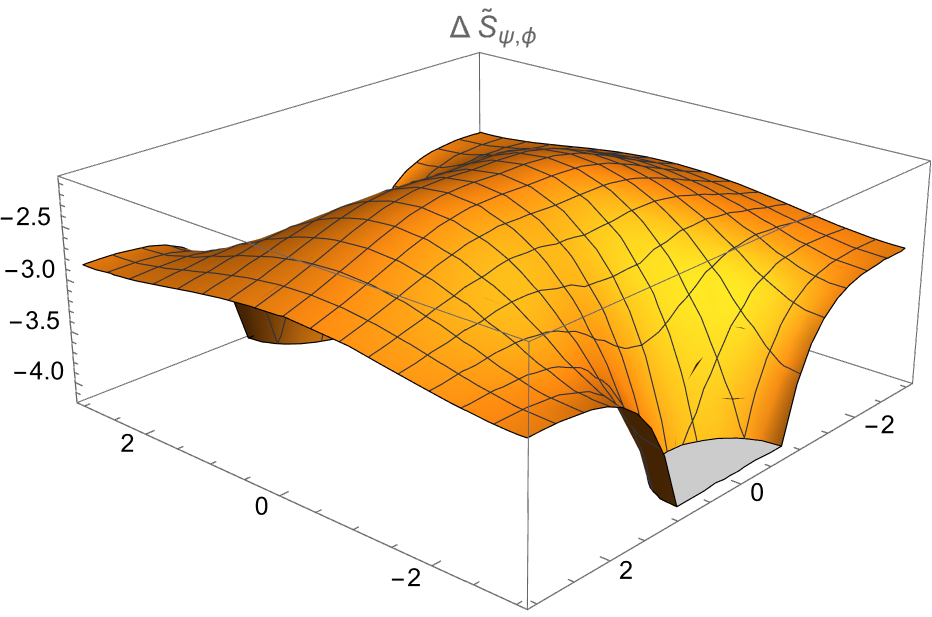}
    \caption{Plot of  $\Delta \tilde{S}_{\psi,\phi}$ as a function of the parameter $\gamma$, where both the real and imaginary parts of $\gamma$ range from $-3$ to $3$.}\label{F2}
  \end{minipage}
\end{figure}

~\\
In Fig.\ref{F1} and Fig.\ref{F2} we show the numerical results of the function $\Delta S_{\psi,\phi}$ and $\Delta \tilde{S}_{\psi,\phi}$. We can see that the function $\Delta \tilde{S}_{\psi,\phi}$ is negative, which supports the upper bound (\ref{upperbound}). While $\Delta S_{\psi,\phi}$ is positive in the same region of $\gamma$.

In the above example, we omitted the term of residue summation as we relied on the findings from section.\ref{sectionqubit}, where we expanded the function as a series of $\gamma$. However, for large $|\gamma|$, this expansion may pose problems. In the exact expression, the EE $S(\rho(\zeta))$ would include a logarithmic term, potentially encountering the branch cut of the logarithmic function. In the subsequent section, we will briefly discuss the branch cut problem. Nonetheless, it appears that the upper bound (\ref{upperbound}) remains correct even for large $|\gamma|$. The upper bound (\ref{upperbound}) is indeed a very loose bound. In the case of the perturbed state example, we observe that the upper bound significantly exceeds $|S(\mathcal{T}_A)|$. Similarly, in the qubit example, this bound appears to be excessively large. It might be plausible to adjust the construction by selecting a different state $|\xi'(\theta)\rangle$ (\ref{xiprime}).

\section{Discussion and conclusion}

In this paper, we explore a novel operator sum rule (\ref{sumrulenew}) applicable to the reduced transition matrix $\transA$ and the reduced density matrix $\rho_A(\theta)$. This rule enables the establishment of a relationship between off-diagonal elements and the diagonal elements (\ref{correlatorsumrule}). Moreover, we establish a connection between the pseudo-Rényi entropy and the Rényi entropy. In the previous work \cite{Guo:2023aio}, we developed a sum rule using discrete Fourier transformation. However, this approach failed to provide a smooth limit for the PE derived from the sum rule for pseudo-Rényi entropy. Through the utilization of our new form of the sum rule, we successfully derive a sum rule for PE (\ref{sumrulepseudo}).

Furthermore, we demonstrate that the transition matrix and the density matrix can be treated uniformly through the analytical continuation of the superposition parameter $\theta$ to a complex variable $\zeta$, as detailed in section.\ref{superpositionsection}. This approach allows for the extraction of the transition matrix $\trans$ via contour integration at infinity on the $\zeta$ plane. By deforming the contour appropriately, we can readily construct the operator sum rule (\ref{sumrulenew}). This analytical continuation becomes indispensable for deriving the complete form of the sum rule for PE. As elucidated in section. \ref{sectionanalytical}, the accurate sum rule for PE necessitates the inclusion of the residue summation term (\ref{PEsumrulefull}). This assertion is validated through examples discussed in subsequent sections.

In section.\ref{sectionapplication}, we delve into the practical implications of the sum rule. One noteworthy application involves gaining insights into the gravity dual of non-Hermitian transition matrices. Our findings reveal that the scaling behavior of connected correlation functions of single trace operators maintains consistency in both the superposition state $|\xi(\theta)\rangle$ and the transition matrix in the large $N$ limit. This observation suggests that a gravity geometry dual should exist for the transition matrix when the superposition state $|\xi(\theta)\rangle$ possesses one. Moreover, the sum rule establishes a connection between two bulk geometries and their on-shell actions. However, as present we lack explicit examples to illustrate this concept. The primary challenge lies in the scarcity of examples demonstrating the exact duality between bulk geometry and boundary states. Recent studies have introduced a class of geometric states known as fixed area states, which are anticipated to exhibit flat entanglement spectra \cite{Dong:2018seb}\cite{Akers:2018fow}, see also \cite{Dong:2019piw}-\cite{Guo:2021tzs}. Additionally, there have been some proposals regarding exact states that are dual to fixed area states in boundary CFTs. These investigations hold promise for providing examples that show the significance of the sum rule within the bulk context.

Another intriguing application lies in determining the upper bound of the absolute value of $S(\transA)$. Utilizing the sum rule for PE, this upper bound becomes associated with the upper limit of a superposition state, a topic previously explored in \cite{superposition}. In our paper, we provide a preliminary bound, which we validate through simple examples. However, we anticipate that the bound could be refined by employing a similar method with a more careful construction of the superposition state $|\xi'(\theta)\rangle$. In the work \cite{Mollabashi:2020yie}, the author proposes that the function (\ref{difference}) is non-positive in QFTs. The sum rule for PE may serve as a valuable tool in either verifying or disproving this proposition.

However, it's crucial to acknowledge that there are still subtleties regarding the sum rule for PE. Constructing the sum rule for PE hinges on investigating the singularities of the function $\mathcal{N}_\zeta S(\rho_A(\zeta))$. The entropy function $S(\rho_A(\zeta))$ typically involves the logarithmic function. In this paper, we haven't considered the branch cut arising from the logarithmic function. In all the examples discussed, we assume that one can expand the results with respect to some small parameter, effectively choosing a branch of the logarithmic function. However, if one were to employ the full form of the function $S(\rho_A(\zeta))$, it would seem necessary to consider potential modifications due to the branch cut. Currently, we have not found a definitive solution to this issue, but it is an problem we plan to explore in the near future. Nevertheless, it's important to emphasize that our result regarding the sum rule for PE is applicable in a wide range of examples.
\\~

~\\
{\bf Acknowledgements}
We would like to thank Xin Gao, Song He, Houwen Wu, Peng Wang, Haitang Yang, Jiaju Zhang, Long Zhao, Yu-Xuan Zhang and Zi-Xuan Zhang for useful discussions. Specially, we would like to thank Jiaju Zhang for early collaboration on this work.
WZG is supposed by the National Natural Science Foundation of China under Grant No.12005070 and the Fundamental Research Funds for the Central Universities under Grants NO.2020kfyXJJS041.

\appendix

\section{Second order perturbation of entropy}\label{appendixperturbation}
 
Given the operator $M$ with $tr M=1$, the entropy can be written as
\bea
S(M)=-\tr M\log M=\int_0^\infty d\beta \left(tr M\frac{1}{\beta+M}-\frac{1}{\beta+1} \right).
\eea
For the perturbed operator $M+\delta M$ we have
\bea
S(M+\delta M)=\int_0^\infty d\beta \left(tr (M+\delta M )\frac{1}{\beta+M+\delta M}-\frac{1}{\beta+1} \right).
\eea
We should use the following equation 
\bea
&&(\beta+M +\delta M)^{-1}\nn \\
&&=[(\beta+M)(1+(\beta +M)^{-1}\delta M)]^{-1}\nn \\
&&=(1+(\beta +M)^{-1}\delta M)^{-1}(\beta+M)^{-1}\nn \\
&&=\left(1-(\beta +M)^{-1}\delta M+(\beta +M)^{-1}\delta M(\beta +M)^{-1}\delta M\right)(\beta +M)^{-1}+O(\delta M^3).\nn
\eea
Now we can find the second order contributions in $S(M+\delta M)$, which is
\bea
&&\delta S^{(2)}=\int_0^\infty d\beta \left(-tr \delta M (\beta +M)^{-1}\delta M (\beta +M)^{-1} \right)\nn \\
&&\phantom{\delta S^{(2)}}+\int_0^\infty d\beta tr \left( M (\beta +M)^{-1}\delta M(\beta +M)^{-1}\delta M(\beta +M)^{-1}\right) \nn \\
&&\phantom{\delta S^{(2)}}=-\int_0^\infty d\beta \beta tr \left((\beta +M)^{-2}\delta M(\beta +M)^{-1}\delta M \right).
\eea

\section{Derivation of the relation (\ref{correlatorsumrule})}

One could derive the relation (\ref{correlatorsumrule}) by utilizing the property of Fourier transformation. By definition, we have
\bea
&&\frac{1}{2\pi }\int_{-\pi}^\pi d\theta e^{-i n\theta}\mathcal{N}_\theta^{n} \prod_{j=1}^{m}\langle \xi(\theta)| \mA_j|\xi(\theta)\rangle\nn \\
&&=\frac{1}{2\pi\langle \phi|\psi\rangle^n}\int_{-\pi}^\pi d\theta e^{-i n\theta} G(e^{i\theta}),
\eea
with 
\bea\label{Four}
G(e^{i\theta})=(2+e^{i\theta}\langle \phi|\psi\rangle+e^{-i\theta}\langle\psi| \phi\rangle)^{n-m} \prod_{j=1}^m (\langle \phi|+e^{-i\theta}\langle \psi|) \mA_j (|\phi\rangle+e^{i\theta}|\psi\rangle).
\eea
The function $G(e^{i\theta})$ can be expanded as polynomials of $e^{i\theta}$ as long as $n,m$ are integers and $n\geq m$. The highest power is of $e^{i n \theta}$ with the coefficient
\bea
\langle \phi|\psi\rangle^{n-m} \prod_{j=1}^m \langle \phi|\mathcal{A}_j|\psi\rangle.
\eea
This is the only surviving term through Fourier transformation in (\ref{Four}). It can be demonstrated that the result of (\ref{Four}) is 
\bea
\prod_{j=1}^{m}\frac{\langle \phi| \mA_j|\psi\rangle}{\langle \phi| \psi\rangle}.
\eea
Thus, the relation (\ref{correlatorsumrule}) is proven. In section.\ref{sectioncorrelatorsumrulealt}, where we develop the sum rule for PE with short interval expansion, we arrive at a formula (\ref{sumrulecorrelatoralt}) akin to (\ref{correlatorsumrule}). If one begins with the sum rule for pseudo-Rényi entropy (\ref{pseudoRenyisumrule}) and employs the short interval expansion, the same formula (\ref{correlatorsumrule}) can be derived.

\begin{thebibliography}{00}
\bibitem{Amico:2007ag}
L.~Amico, R.~Fazio, A.~Osterloh and V.~Vedral, \textit{{Entanglement in
  many-body systems}}, \href{http://dx.doi.org/10.1103/RevModPhys.80.517}{Rev.
  Mod. Phys. {\bfseries 80}, 517 (2008)},
  [\href{https://arxiv.org/abs/quant-ph/0703044}{{\ttfamily
  arXiv:quant-ph/0703044}}].

\bibitem{Eisert:2008ur}
J.~Eisert, M.~Cramer and M.~B. Plenio, \textit{{Area laws for the entanglement
  entropy - a review}}, \href{http://dx.doi.org/10.1103/RevModPhys.82.277}{Rev.
  Mod. Phys. {\bfseries 82}, 277--306 (2010)},
  [\href{https://arxiv.org/abs/0808.3773}{{\ttfamily arXiv:0808.3773}}].

\bibitem{Calabrese:2009qy}
P.~Calabrese and J.~Cardy, \textit{{Entanglement entropy and conformal field
  theory}}, \href{http://dx.doi.org/10.1088/1751-8113/42/50/504005}{J. Phys. A:
  Math. Gen. {\bfseries 42}, 504005 (2009)},
  [\href{https://arxiv.org/abs/0905.4013}{{\ttfamily arXiv:0905.4013}}].

\bibitem{Rangamani:2016dms}
M.~Rangamani and T.~Takayanagi, \textit{{Holographic Entanglement Entropy}},
  \href{http://dx.doi.org/10.1007/978-3-319-52573-0}{Lect. Notes Phys.
  {\bfseries 931}, 1--246 (2017)},
  [\href{https://arxiv.org/abs/1609.01287}{{\ttfamily arXiv:1609.01287}}].

\bibitem{Ryu:2006bv}
S.~Ryu and T.~Takayanagi, \textit{{Holographic Derivation of Entanglement
  Entropy from the anti-de Sitter Space/Conformal Field Theory
  Correspondence}},
  \href{http://dx.doi.org/10.1103/PhysRevLett.96.181602}{Phys. Rev. Lett.
  {\bfseries 96}, 181602 (2006)},
  [\href{https://arxiv.org/abs/hep-th/0603001}{{\ttfamily
  arXiv:hep-th/0603001}}].

\bibitem{Hubeny:2007xt}
V.~E. Hubeny, M.~Rangamani and T.~Takayanagi, \textit{{A Covariant holographic
  entanglement entropy proposal}},
  \href{http://dx.doi.org/10.1088/1126-6708/2007/07/062}{JHEP {\bfseries 07}
  (2007) 062}, [\href{https://arxiv.org/abs/0705.0016}{{\ttfamily
  arXiv:0705.0016}}].



\bibitem{VanRaamsdonk:2010pw}
M.~Van~Raamsdonk, \textit{{Building up spacetime with quantum entanglement}},
  \href{http://dx.doi.org/10.1007/s10714-010-1034-0}{Gen. Rel. Grav. {\bfseries
  42}, 2323 (2010)}, [\href{https://arxiv.org/abs/1005.3035}{{\ttfamily
  arXiv:1005.3035}}]. [\href{http://dx.doi.org/10.1142/S0218271810018529}
  {\textit{Int. J. Mod. Phys.} {\bfseries D19} (2010) 2429}].
\bibitem{Wall:2012uf}
  A.~C.~Wall, \textit{Maximin Surfaces, and the Strong Subadditivity of the Covariant Holographic Entanglement Entropy},
  \href{http://dx.doi.org/10.1088/0264-9381/31/22/225007}{Class.\ Quant.\ Grav.\  {\bf 31}, no. 22, 225007 (2014)}, [\href{https://arxiv.org/abs/1211.3494}{{\ttfamily arXiv:1211.3494[hep-th]}}].
  
\bibitem{Almheiri:2014lwa}
A.~Almheiri, X.~Dong and D.~Harlow, \textit{{Bulk Locality and Quantum Error
  Correction in AdS/CFT}},
  \href{http://dx.doi.org/10.1007/JHEP04(2015)163}{JHEP {\bfseries 04} (2015)
  163}, [\href{https://arxiv.org/abs/1411.7041}{{\ttfamily arXiv:1411.7041}}].
\bibitem{Dong:2016eik}
  X.~Dong, D.~Harlow and A.~C.~Wall,
  \textit{Reconstruction of Bulk Operators within the Entanglement Wedge in Gauge-Gravity Duality}, \href{http://dx.doi.org/10.1103/PhysRevLett.117.021601}{Phys.\ Rev.\ Lett.\  {\bf 117}, no. 2, 021601 (2016)}, [\href{https://arxiv.org/abs/1601.05416}{{\ttfamily arXiv:1601.05416[hep-th]}}].

\bibitem{Penington:2019npb}
G.~Penington, \textit{{Entanglement Wedge Reconstruction and the Information
  Paradox}}, \href{http://dx.doi.org/10.1007/JHEP09(2020)002}{JHEP {\bfseries
  09} (2020) 002}, [\href{https://arxiv.org/abs/1905.08255}{{\ttfamily
  arXiv:1905.08255}}].

\bibitem{Almheiri:2019psf}
A.~Almheiri, N.~Engelhardt, D.~Marolf and H.~Maxfield, \textit{{The entropy of
  bulk quantum fields and the entanglement wedge of an evaporating black
  hole}}, \href{http://dx.doi.org/10.1007/JHEP12(2019)063}{JHEP {\bfseries 12}
  (2019) 063}, [\href{https://arxiv.org/abs/1905.08762}{{\ttfamily
  arXiv:1905.08762}}].











\bibitem{Strominger:2001pn}
A.~Strominger, \textit{The dS / CFT correspondence},
\href{http://dx.doi.org/10.1088/1126-6708/2001/10/034}{JHEP \textbf{10} (2001), 034}, [\href{https://arxiv.org/abs/hep-th/0106113}{{\ttfamily
  arXiv:0106113[hep-th]}}].

\bibitem{Maldacena:2002vr}
J.~M.~Maldacena, \textit{Non-Gaussian features of primordial fluctuations in single field inflationary models}, \href{http://dx.doi.org/10.1088/1126-6708/2003/05/013}{JHEP \textbf{05} (2003), 013}, [\href{https://arxiv.org/abs/astro-ph/0210603}{{\ttfamily
  astro-ph/0210603 [astro-ph]}}].



\bibitem{Doi:2022iyj}
K.~Doi, J.~Harper, A.~Mollabashi, T.~Takayanagi and Y.~Taki,
  \textit{{Pseudoentropy in dS/CFT and Timelike Entanglement Entropy}},
  \href{http://dx.doi.org/10.1103/PhysRevLett.130.031601}{Phys. Rev. Lett.
  {\bfseries 130}, 031601 (2023)},
  [\href{https://arxiv.org/abs/2210.09457}{{\ttfamily arXiv:2210.09457}}].

\bibitem{Nakata:2020luh}
Y.~Nakata, T.~Takayanagi, Y.~Taki, K.~Tamaoka and Z.~Wei, \textit{{New
  holographic generalization of entanglement entropy}},
  \href{http://dx.doi.org/10.1103/PhysRevD.103.026005}{Phys. Rev. D {\bfseries
  103}, 026005 (2021)}, [\href{https://arxiv.org/abs/2005.13801}{{\ttfamily
  arXiv:2005.13801}}].

\bibitem{Murciano:2021dga}
S.~Murciano, P.~Calabrese and R.~M. Konik, \textit{{Generalized entanglement
  entropies in two-dimensional conformal field theory}},
  \href{http://dx.doi.org/10.1007/JHEP05(2022)152}{JHEP {\bfseries 05} (2022)
  152}, [\href{https://arxiv.org/abs/2112.09000}{{\ttfamily
  arXiv:2112.09000}}].
\bibitem{Mollabashi:2020yie}
A.~Mollabashi, N.~Shiba, T.~Takayanagi, K.~Tamaoka and Z.~Wei, \textit{{Pseudo
  Entropy in Free Quantum Field Theories}},
  \href{http://dx.doi.org/10.1103/PhysRevLett.126.081601}{Phys. Rev. Lett.
  {\bfseries 126}, 081601 (2021)},
  [\href{https://arxiv.org/abs/2011.09648}{{\ttfamily arXiv:2011.09648}}].


\bibitem{Guo:2022jzs}
W.-z. Guo, S.~He and Y.-X. Zhang, \textit{{Constructible reality condition of
  pseudoentropy via pseudo-Hermiticity}},
  \href{http://dx.doi.org/10.1007/JHEP05(2023)021}{JHEP {\bfseries 05} (2023)
  021}, [\href{https://arxiv.org/abs/2209.07308}{{\ttfamily
  arXiv:2209.07308}}].





\bibitem{Mollabashi:2021xsd}
A.~Mollabashi, N.~Shiba, T.~Takayanagi, K.~Tamaoka and Z.~Wei, \textit{{Aspects
  of pseudoentropy in field theories}},
  \href{http://dx.doi.org/10.1103/PhysRevResearch.3.033254}{Phys. Rev. Res.
  {\bfseries 3}, 033254 (2021)},
  [\href{https://arxiv.org/abs/2106.03118}{{\ttfamily arXiv:2106.03118}}].

\bibitem{Nishioka:2021cxe}
T.~Nishioka, T.~Takayanagi and Y.~Taki, \textit{{Topological pseudoentropy}},
  \href{http://dx.doi.org/10.1007/JHEP09(2021)015}{JHEP {\bfseries 09} (2021)
  015}, [\href{https://arxiv.org/abs/2107.01797}{{\ttfamily
  arXiv:2107.01797}}].

\bibitem{Goto:2021kln}
K.~Goto, M.~Nozaki and K.~Tamaoka, \textit{{Subregion spectrum form factor via
  pseudoentropy}}, \href{http://dx.doi.org/10.1103/PhysRevD.104.L121902}{Phys.
  Rev. D {\bfseries 104}, L121902 (2021)},
  [\href{https://arxiv.org/abs/2109.00372}{{\ttfamily arXiv:2109.00372}}].

\bibitem{Akal:2021dqt}
I.~Akal, T.~Kawamoto, S.-M. Ruan, T.~Takayanagi and Z.~Wei, \textit{{Page curve
  under final state projection}},
  \href{http://dx.doi.org/10.1103/PhysRevD.105.126026}{Phys. Rev. D {\bfseries
  105}, 126026 (2022)}, [\href{https://arxiv.org/abs/2112.08433}{{\ttfamily
  arXiv:2112.08433}}].

\bibitem{Miyaji:2021lcq}
M.~Miyaji,
\textit{Island for gravitationally prepared state and pseudo entanglement wedge},
\href{http://dx.doi.org/10.1007/JHEP12(2021)013}{
JHEP \textbf{12}, 013 (2021)},
[\href{https://arxiv.org/abs/2109.03830}{{\ttfamily arXiv:2109.03830}}].



\bibitem{Guo:2022sfl}
W.-z. Guo, S.~He and Y.-X. Zhang, \textit{{On the real-time evolution of
  pseudo-entropy in 2d CFTs}},
  \href{http://dx.doi.org/10.1007/JHEP09(2022)094}{JHEP {\bfseries 09} (2022)
  094}, [\href{https://arxiv.org/abs/2206.11818}{{\ttfamily arXiv:2206.11818}}].

\bibitem{Ishiyama:2022odv}
Y.~Ishiyama, R.~Kojima, S.~Matsui and K.~Tamaoka,
\textit{Notes on pseudoentropy amplification},
\href{http://dx.doi.org/10.1093/ptep/ptac112}{PTEP \textbf{2022}, no.9, 093B10 (2022)},
[\href{https://arxiv.org/abs/2206.14551}{{\ttfamily arXiv:2206.14551}}]
\bibitem{Mukherjee:2022jac}
J.~Mukherjee,
\textit{Pseudo Entropy in U(1) gauge theory},
\href{https://link.springer.com/article/10.1007/JHEP10(2022)016}{JHEP \textbf{10}(2022)016},
[\href{https://arxiv.org/abs/2205.08179}{\ttfamily{arXiv:2205.08179}}].

\bibitem{Li:2022tsv}
Z.~Li, Z.-Q. Xiao and R.-Q. Yang, \textit{{On holographic time-like
  entanglement entropy}}, \href{http://dx.doi.org/10.1007/JHEP04(2023)004}{JHEP
  {\bfseries 04} (2023) 004},
  [\href{https://arxiv.org/abs/2211.14883}{{\ttfamily arXiv:2211.14883}}].

\bibitem{He:2023eap}
S.~He, J.~Yang, Y.-X. Zhang and Z.-X. Zhao, \textit{{Pseudo-entropy for
  descendant operators in two-dimensional conformal field theories}},
  \href{https://arxiv.org/abs/2301.04891}{{\ttfamily arXiv:2301.04891}}.

\bibitem{Narayan:2022afv}
K.~Narayan, \textit{{de Sitter space, extremal surfaces, and time
  entanglement}}, \href{http://dx.doi.org/10.1103/PhysRevD.107.126004}{Phys.
  Rev. D {\bfseries 107}, 126004 (2023)},
  [\href{https://arxiv.org/abs/2210.12963}{{\ttfamily arXiv:2210.12963}}].

\bibitem{Doi:2023zaf}
K.~Doi, J.~Harper, A.~Mollabashi, T.~Takayanagi and Y.~Taki, \textit{{Timelike
  entanglement entropy}}, \href{http://dx.doi.org/10.1007/JHEP05(2023)052}{JHEP
  {\bfseries 05} (2023) 052},
  [\href{https://arxiv.org/abs/2302.11695}{{\ttfamily arXiv:2302.11695}}].

\bibitem{Kawamoto:2023nki}
T.~Kawamoto, S.~M.~Ruan, Y.~k.~Suzuki and T.~Takayanagi,
\textit{A half de Sitter holography},
\href{https://link.springer.com/article/10.1007/JHEP10(2023)137}{JHEP \textbf{10} (2023) 137},
[\href{https://arxiv.org/abs/2306.07575}{\ttfamily{arXiv:2306.07575}}].

\bibitem{Narayan:2023ebn}
K.~Narayan and H.~K. Saini, \textit{{Notes on time entanglement and
  pseudo-entropy}},  \href{https://arxiv.org/abs/2303.01307}{{\ttfamily
  arXiv:2303.01307}}.

\bibitem{He:2023wko}
S.~He, J.~Yang, Y.~X.~Zhang and Z.~X.~Zhao,
\textit{Pseudo entropy of primary operators in $T\bar{T}$/$J\bar{T}$-deformed CFTs},
\href{https://arxiv.org/abs/2305.10984}{{\ttfamily arXiv:2305.10984}}

\bibitem{Jiang:2023ffu}
X.~Jiang, P.~Wang, H.~Wu and H.~Yang, \textit{{Timelike entanglement entropy
  and $T\bar{T}$ deformation}},
  \href{https://arxiv.org/abs/2302.13872}{{\ttfamily arXiv:2302.13872}}.

\bibitem{Parzygnat:2023avh}
A.~J.~Parzygnat, T.~Takayanagi, Y.~Taki and Z.~Wei, \textit{SVD entanglement entropy}, 
\href{https://link.springer.com/article/10.1007/JHEP12(2023)123}{JHEP \textbf{12} (2023), 123},
[\href{https://arxiv.org/abs/2307.06531}{\ttfamily{arXiv:2307.06531}}].


\bibitem{Guo:2023aio}
W.~z.~Guo and J.~Zhang,
\textit{Sum rule for pseudo-R\'enyi entropy}, \href{https://journals.aps.org/prd/abstract/10.1103/PhysRevD.109.106008}{Phys. Rev. D 109, 106008(2024)},
\href{https://arxiv.org/abs/2308.05261}{{\ttfamily arXiv:2308.05261}}.

\bibitem{He:2023syy}
S.~He, Y.~X.~Zhang, L.~Zhao and Z.~X.~Zhao,
\textit{Entanglement and Pseudo Entanglement Dynamics versus Fusion in CFT},
\href{https://arxiv.org/abs/2312.02679}{{\ttfamily arXiv:2312.02679}}.

\bibitem{Das:2023yyl}
A.~Das, S.~Sachdeva and D.~Sarkar,
\textit{Bulk reconstruction using timelike entanglement in (A)dS},
\href{https://arxiv.org/abs/2312.16056}{\ttfamily{arXiv:2312.16056 }}.

\bibitem{Chu:2023zah}
C.~S.~Chu and H.~Parihar,
\textit{Time-like entanglement entropy in AdS/BCFT},
\href{https://link.springer.com/article/10.1007/JHEP06(2023)173}{
JHEP \textbf{06} (2023) 173},
[\href{https://arxiv.org/abs/2304.10907}{\ttfamily {arXiv:2304.10907}}].

\bibitem{Jiang:2023loq}
X.~Jiang, P.~Wang, H.~Wu and H.~Yang,
\textit{Timelike entanglement entropy in dS$_{3}$/CFT$_{2}$},
\href{https://link.springer.com/article/10.1007/JHEP08(2023)216}{JHEP \textbf{08} (2023) 216},
[\href{https://arxiv.org/abs/2304.10376}{\ttfamily{arXiv:2304.10376}}].

\bibitem{Chen:2023gnh}
Z.~Chen,
\textit{Complex-valued Holographic Pseudo Entropy via Real-time AdS/CFT Correspondence},
\href{https://arxiv.org/abs/2302.14303}{\ttfamily{arXiv:2302.14303}}.

\bibitem{He:2023ubi}
P.~Z.~He and H.~Q.~Zhang,
\textit{Timelike Entanglement Entropy from Rindler Method},
\href{https://arxiv.org/abs/2307.09803}{\ttfamily{arXiv:2307.09803}}.
\bibitem{Omidi:2023env}
F.~Omidi,
\textit{Pseudo R\'enyi Entanglement Entropies For an Excited State and Its Time Evolution in a 2D CFT},
\href{https://arxiv.org/abs/2309.04112}{\ttfamily{arXiv:2309.04112}}.

\bibitem{Narayan:2023zen}
K.~Narayan,
\textit{Further remarks on de Sitter space, extremal surfaces and time entanglement},
\href{https://arxiv.org/abs/2310.00320}{\ttfamily{arXiv:2310.00320}}.


\bibitem{Guo:2023tjv}
W.~z.~Guo, Y.~z.~Jiang, \textit{Pseudo entropy and pseudo-Hermiticity in quantum field theories}, \href{https://link.springer.com/article/10.1007/JHEP05(2024)071}{JHEP \textbf{05} (2024), 071}, \href{https://arxiv.org/abs/2311.01045}{\ttfamily{arXiv:2311.01045}}.



\bibitem{Kanda:2023jyi}
H.~Kanda, T.~Kawamoto, Y.~k.~Suzuki, T.~Takayanagi, K.~Tasuki and Z.~Wei,
\textit{Entanglement Phase Transition in Holographic Pseudo Entropy},
\href{https://arxiv.org/abs/2311.13201}{\ttfamily{arXiv:2311.13201}}.

\bibitem{Shinmyo:2023eci}
K.~Shinmyo, T.~Takayanagi and K.~Tasuki,
\textit{Pseudo entropy under joining local quenches},
\href{https://arxiv.org/abs/2310.12542}{\ttfamily{arXiv:2310.12542}}.



\bibitem{Basak:2023otu}
J.~K.~Basak, A.~Chakraborty, C.~S.~Chu, D.~Giataganas and H.~Parihar,
\textit{Massless Lifshitz field theory for arbitrary z},
\href{https://link.springer.com/article/10.1007/JHEP05(2024)284}{JHEP \textbf{05} (2024), 284}, \href{https://arxiv.org/abs/2312.16284}{\ttfamily{arXiv:2312.16284}}.

\bibitem{Afrasiar:2024lsi}
M.~Afrasiar, J.~K.~Basak and D.~Giataganas,
\textit{Timelike Entanglement Entropy and Phase Transitions in non-Conformal Theories},
\href{https://arxiv.org/abs/2404.01393}{\ttfamily{arXiv:2404.01393}}.




\bibitem{Guo:2024lrr}
W.~z.~Guo, S.~He and Y.~X.~Zhang, \textit{Relation between timelike and spacelike entanglement entropy}, [\href{https://arxiv.org/abs/2402.00268}{{\ttfamily
  arXiv:2402.00268 }}].

\bibitem{He:2024jog}
S.~He, P.~H.~C.~Lau and L.~Zhao, \textit{Detecting quantum chaos via pseudo-entropy and negativity},
[\href{https://arxiv.org/abs/2403.05875}{{\ttfamily
  arXiv:2403.05875 }}].

 
\bibitem{Lashkari:2014yva}
N.~Lashkari, \textit{Relative Entropies in Conformal Field Theory},
\href{http://dx.doi.org/10.1103/PhysRevLett.113.051602}{Phys. Rev. Lett. \textbf{113} (2014), 051602},
  [\href{https://arxiv.org/abs/1404.3216}{{\ttfamily arXiv:1404.3216}}].

  
  
 
  
\bibitem{DHoker:2020bcv}
E.~D'Hoker, X.~Dong and C.~H.~Wu, \textit{An alternative method for extracting the von Neumann entropy from R\'enyi entropies}, \href{http://dx.doi.org/10.1007/JHEP01(2021)042}{JHEP \textbf{01} (2021), 042},
  [\href{https://arxiv.org/abs/2008.10076}{{\ttfamily arXiv:2008.10076}}].





\bibitem{Cardy:2007mb}
J.~L. Cardy, O.~A. Castro-Alvaredo and B.~Doyon, \textit{{Form factors of
  branch-point twist fields in quantum integrable models and entanglement
  entropy}}, \href{http://dx.doi.org/10.1007/s10955-007-9422-x}{J. Stat. Phys.
  {\bfseries 130}, 129 (2008)},
  [\href{https://arxiv.org/abs/0706.3384}{{\ttfamily arXiv:0706.3384}}].

\bibitem{Headrick:2010zt}
M.~Headrick, \textit{{Entanglement R\'enyi entropies in holographic theories}},
  \href{http://dx.doi.org/10.1103/PhysRevD.82.126010}{Phys. Rev. D {\bfseries
  82}, 126010 (2010)}, [\href{https://arxiv.org/abs/1006.0047}{{\ttfamily
  arXiv:1006.0047}}].

\bibitem{Calabrese:2010he}
P.~Calabrese, J.~Cardy and E.~Tonni, \textit{{Entanglement entropy of two
  disjoint intervals in conformal field theory II}},
  \href{http://dx.doi.org/10.1088/1742-5468/2011/01/P01021}{J. Stat. Mech.
  (2011) P01021}, [\href{https://arxiv.org/abs/1011.5482}{{\ttfamily
  arXiv:1011.5482}}].

\bibitem{Rajabpour:2011pt}
M.~A. Rajabpour and F.~Gliozzi, \textit{{Entanglement Entropy of Two Disjoint
  Intervals from Fusion Algebra of Twist Fields}},
  \href{http://dx.doi.org/10.1088/1742-5468/2012/02/P02016}{J. Stat. Mech.
  (2012) P02016}, [\href{https://arxiv.org/abs/1112.1225}{{\ttfamily
  arXiv:1112.1225}}].

\bibitem{Chen:2013kpa}
B.~Chen and J.-j. Zhang, \textit{{On short interval expansion of R\'enyi
  entropy}}, \href{http://dx.doi.org/10.1007/JHEP11(2013)164}{JHEP {\bfseries
  11} (2013) 164}, [\href{https://arxiv.org/abs/1309.5453}{{\ttfamily
  arXiv:1309.5453}}].





\bibitem{Aharony:1999ti}
O.~Aharony, S.~S.~Gubser, J.~M.~Maldacena, H.~Ooguri and Y.~Oz,
\textit{Large N field theories, string theory and gravity},
\href{http://dx.doi.org/10.1016/S0370-1573(99)00083-6}
{Phys. Rept. \textbf{323} (2000), 183-386}, [\href{https://arxiv.org/abs/hep-th/9905111}{{\ttfamily
 [arXiv:hep-th/9905111 ]}}].


\bibitem{Harlow:2018fse}
D.~Harlow, \textit{{TASI Lectures on the Emergence of Bulk Physics in
  AdS/CFT}}, \href{http://dx.doi.org/10.22323/1.305.0002}{PoS {\bfseries
  TASI2017}, 002 (2018)}, [\href{https://arxiv.org/abs/1802.01040}{{\ttfamily
  arXiv:1802.01040}}].

\bibitem{Dong:2016fnf}
X.~Dong, \textit{{The gravity dual of R\'enyi entropy}},
  \href{http://dx.doi.org/10.1038/ncomms12472}{Nature Commun. {\bfseries 7},
  12472 (2016)}, [\href{https://arxiv.org/abs/1601.06788}{{\ttfamily
  arXiv:1601.06788}}].

  
\bibitem{superposition}
N. Linden, S. Popescu, and J. A. Smolin,  \textit{Entanglement of Superpositions}, \href{https://doi.org/10.1103/PhysRevLett.97.100502}{Phys. Rev. Lett.
97, 100502 (2006)}.






\bibitem{Dong:2018seb}
X.~Dong, D.~Harlow and D.~Marolf, \textit{Flat entanglement spectra in fixed-area states of quantum gravity}, \href{http://dx.doi.org/10.1007/JHEP10(2019)240}{JHEP \textbf{10}, 240 (2019)}, [\href{https://arxiv.org/abs/1811.05382}{{\ttfamily
  arXiv:1811.05382 [hep-th]}}].







\bibitem{Akers:2018fow}
C.~Akers and P.~Rath, \textit{Holographic Renyi Entropy from Quantum Error Correction},
\href{http://dx.doi.org/10.1007/JHEP05(2019)052}{JHEP \textbf{05}, 052 (2019)}, [\href{https://arxiv.org/abs/1811.05171}{{\ttfamily
  arXiv:1811.05171 [hep-th]}}].


\bibitem{Dong:2019piw}
X.~Dong and D.~Marolf, \textit{One-loop universality of holographic codes}, \href{http://dx.doi.org/10.1007/JHEP05(2019)052}{JHEP \textbf{03}, 191 (2020)}, [\href{https://arxiv.org/abs/1910.06329}{{\ttfamily
  arXiv:1910.06329 [hep-th]}}].

\bibitem{Dong:2020iod}
X.~Dong and H.~Wang, \textit{Enhanced corrections near holographic entanglement transitions: a chaotic case study}, \href{http://dx.doi.org/10.1007/JHEP05(2019)052}{JHEP \textbf{11}, 007 (2020)}, [\href{https://arxiv.org/abs/2006.10051}{{\ttfamily
  arXiv:2006.10051 [hep-th]}}].

\bibitem{Marolf:2020vsi}
D.~Marolf, S.~Wang and Z.~Wang, \textit{Probing phase transitions of holographic entanglement entropy with fixed area states}, \href{http://dx.doi.org/10.1007/JHEP12(2020)084}{JHEP \textbf{12}, 084 (2020)}, [\href{https://arxiv.org/abs/2006.10089}{{\ttfamily
 arXiv:2006.10089 [hep-th]}}].


\bibitem{Akers:2020pmf}
C.~Akers and G.~Penington, \textit{Leading order corrections to the quantum extremal surface prescription}, \href{http://dx.doi.org/10.1007/JHEP04(2021)062}{JHEP \textbf{04}, 062 (2021)}, [\href{
https://doi.org/10.48550/arXiv.2008.03319
}{{\ttfamily
arXiv:2008.03319 [hep-th]}}].



\bibitem{Dong:2021clv}
X.~Dong, X.~L.~Qi and M.~Walter, \textit{Holographic entanglement negativity and replica symmetry breaking}, \href{http://dx.doi.org/10.1007/JHEP06(2021)024}{JHEP \textbf{06}, 024 (2021)}, [\href{https://arxiv.org/abs/2006.10089}{{\ttfamily
arXiv:2101.11029 [hep-th]}}].

\bibitem{Guo:2021tzs}
W.~z.~Guo, \textit{Area operator and fixed area states in conformal field theories},
\href{http://dx.doi.org/10.1103/PhysRevD.106.L061903}{Phys. Rev. D \textbf{106} (2022) no.6, L061903}, [\href{https://arxiv.org/abs/2108.03346}{{\ttfamily
arXiv:2108.03346 [hep-th]}}].


  
\end{thebibliography}

\end{document}